\documentclass{emulateapj}
\shorttitle{Constraints from classicals}
\shortauthors{Dawson and Murray-Clay}
\newcommand{\gkbo}{g_{\rm KBO}}
\newcommand{\aN}{a_{\rm N}}
\newcommand{\eN}{e_{\rm N}}

\newcommand{\varpiN}{\varpi_{\rm N}}
\newcommand{\dotvarpiN}{\dot{\varpi}_{\rm N}}
\newcommand{\mN}{m_{\rm N}}
\begin{document}
\title{Neptune's wild days: constraints from the eccentricity distribution of the classical Kuiper Belt}

\slugcomment{ApJ, 750, 43; submitted: May 25th, 2011; accepted: February 25, 2012}

\author{
Rebekah I. Dawson\altaffilmark{1}
 \&
Ruth Murray-Clay
}
\affil{Harvard-Smithsonian Center for Astrophysics \\ 
60 Garden St, MS-51, Cambridge, MA 02138}

\altaffiltext{1}{{\tt  rdawson@cfa.harvard.edu}}

\begin{abstract}
Neptune's dynamical history shaped the current orbits of Kuiper Belt objects (KBOs), leaving clues to the planet's orbital evolution. In the ``classical" region, a population of dynamically ``hot" high-inclination KBOs overlies a flat ``cold" population with distinct physical properties. Simulations of qualitatively different histories for Neptune Ðincluding smooth migration on a circular orbit or scattering by other planets to a high eccentricity Ð have not simultaneously produced both populations. We explore a general Kuiper Belt assembly model that forms hot classical KBOs interior to Neptune and delivers them to the classical region, where the cold population forms in situ. First, we present evidence that the cold population is confined to eccentricities well below the limit dictated by long-term survival. Therefore Neptune must deliver hot KBOs into the long-term survival region without excessively exciting the eccentricities of the cold population. Imposing this constraint, we explore the parameter space of Neptune's eccentricity and eccentricity damping, migration, and apsidal precession. We rule out much of parameter space, except where Neptune is scattered to a moderately eccentric orbit (e$>$0.15) and subsequently migrates a distance $\Delta\aN=1-6$ AU. Neptune's moderate eccentricity must either damp quickly or be accompanied by fast apsidal precession.  We find that Neptune's high eccentricity alone does not generate a chaotic sea in the classical region. Chaos can result from Neptune's interactions with Uranus, exciting the cold KBOs and placing additional constraints.  Finally, we discuss how to interpret our constraints in the context of the full, complex dynamical history of the solar system.
\end{abstract}


\section{Introduction}

Neptune, with its nearly circular and equatorial orbit, may seem straight-laced compared to the oblique, hot, eccentric, and resonant planets in the extra-solar menagerie. But the highly inclined and eccentric orbits of Pluto \citep{1993M, 1995M} and subsequently discovered Kuiper Belt objects (KBOs) imply that Neptune may have experienced its own ``wild days" in the early solar system. During these wild days, Neptune sculpted the KBOs into four main dynamical classes: objects in mean motion orbital resonance with Neptune (the ``resonant" population), objects that are currently scattering off Neptune (the ``scattering" population), and two populations of ``classical" objects that are currently decoupled from Neptune. One population of classical objects is dynamically ``cold," on nearly circular orbits at low inclinations, and the other classical population is dynamically ``hot" with a range of eccentricities and inclinations. The cold classicals have distinct physical properties from the hot classicals, including colors \citep{2000T,2002T,2008P}, sizes \citep{2001L,2010F}, albedos \citep{2009B}, and binary fraction \citep{2006S,2008N}. A major problem in understanding the formation of the solar system is that, as we will review below, no model of Neptune's dynamical history adequately produces the superposition of hot and cold classicals or accounts for the difference in their physical properties. 

Two types of dynamical sculpting models have been developed to explain, in particular, the population of resonant KBOs. Extensive migration models \citep{1993M,1995M,2005H} propose that Neptune migrated outward by 7-10 AU on a nearly circular orbit from its location of formation, capturing objects into resonance as its resonance locations slowly swept through the Kuiper Belt. This type of model generates the resonant and scattering objects and the cold population (unexcited objects that, in this model, formed in situ) but not the hot population. It also does not match the observed inclination distribution within the resonances. Chaotic capture models \citep{2008L}, inspired by the Nice model \citep[see][and references therein]{2008M}, propose that Neptune was scattered onto a highly eccentric orbit by other planets during a period of instability \citep{1999T}. Neptune's high eccentricity created a chaotic zone in what is now the classical region, and some objects were caught in resonances when Neptune's eccentricity damped. Chaotic capture models produce a hot population: objects that formed in the inner disk, subsequently were scattered by Neptune into the classical region, and then decoupled when Neptune's eccentricity damped. These models also produce a resonant population and scattering population. Although some of the objects delivered into the classical region end up on low-eccentricity orbits, we point out that a cold population confined to low eccentricities is not produced.  Other variations of the Nice model \citep[e.g.][]{2008M} include an in situ population of cold objects, but, over the course of Neptune's evolution, these objects become excited to higher eccentricities. K. Batygin (2010, private communication\footnote{After the submission of this manuscript, \citet{2011B} presented a model in which Neptune underwent a period of high eccentricity and, due to its fast apsidal precession, could avoid disrupting the cold classicals. Because this paper appeared after the submission of our manuscript, we leave a detailed discussion of its results for future work. However, we note that in the particular simulations they presented, the cold classicals are dynamically excited, inconsistent with the constraints we will establish. In Section \ref{subsec:precess}, we explore under what circumstances, if any, this could be avoided.}) has suggested that fast apsidal precession of Neptune's orbit could prevent Neptune from disrupting the cold classicals during its proposed high-eccentricity period, but it remains to be explored under what circumstances this mechanism would work and how it would affect the hot classicals. 

With neither the extensive migration models nor chaotic capture models producing both the hot and cold classicals, the qualitative picture of what happened in the early solar system, including the roles of planet-planet scattering and planetary migration, remains up for debate. It remains a question whether Neptune migrated outward by many AU on a nearly circular orbit, was launched onto an eccentric orbit near its current location, or none of the above. Pinning down Neptune's dynamical history, which should be possible given the constraints from over 500 KBOs with well characterized orbits, would reveal much about the history of our own solar system and about the processes of scattering and migration that shape the architecture of many planetary systems.

Previous models attempted to produce all four dynamical classes of KBOs with 0.5-4 Gyr simulations that included all four giant planets and thousands of massless KBOs \citep[e.g.][]{2005H, 2008L}. It has not been computationally feasible to fully explore parameter space with such extensive simulations. Thus it is unclear whether the dynamical history described by a particular model (1) has trouble producing both the cold and hot populations because there is a qualitative problem with the scenario or, alternatively, because the parameters need to be slightly adjusted; and (2) is unique, or whether another, qualitatively different dynamical history would match the observations just as well. 

Inspired by previous models, we explore a generalization in which Neptune undergoes all potential combinations of high eccentricity, migration, and/or apsidal precession: ``Neptune's wild days." In this generalization, the cold objects form in situ where we observe them today and the hot objects are delivered from the inner disk and superimposed on the cold objects. ``Two-origin" models superimposing a hot classical population from the inner disk on top of a cold population firmed in situ \citep[e.g.][]{2001L,2003G,2008M} have the advantage of explaining the different physical properties of the hot and cold classicals that were discussed above. The different colors, sizes, and albedos of the two populations are accounted for by their formation in different regions of the solar system's proto-planteary disk under different conditions. For instance, chemical differences may result in different colors for objects formed in the inner versus the outer disk \citep{2011BSF}. The cold classicals have a higher binary fraction because any hot classical binaries were likely to have been disrupted when they were scattered from the inner disk to the classical region \citep{2010P} and because binary capture may have been less efficient in the inner disk \citep{2011MS}. However, to date this class of model has not yet been demonstrated to work quantitatively. We consider a generalized two-origin model in which hot classical deliver echoers as a result of scattering by Neptune (rather than due to resonance sweeping as in \citealt{2003G}). Focusing on the consistency of this class of model with the eccentricity distribution of classical KBOs --- unaccounted for by previous model realizations --- we explore the parameter space for this generalized model using several alternative tactics:

\begin{itemize}
\item Instead of attempting to produce a single model, we fully explore the parameter space of Neptune's eccentricity, semimajor axis, migration rate, eccentricity damping rate, and precession rate to assess the consistency of a collection of dynamical histories with the observations. This approach is general in the sense that previous models \citep[e.g.][]{1995M,2008L} are under consideration (corresponding to a particular set of parameters), as well as other regions of parameter space that have not been explicitly considered. We will explore Neptune's inclination and inclination damping rate in a paper currently in preparation (R.I. Dawson and R. Murray-Clay 2012, in preparation). In Section \ref{subsec:combointerpret}, we clarify how to interpret complex solar system histories in the context of this general model.
\item Instead of matching the observations in detail, we focus on matching major qualitative features of the classical KBO eccentricity distribution that are unaffected by observational bias or by the long-term evolution of the solar system (i.e. the evolution that happens over the $\sim$ 4 Gyr after the planets reach their final configuration). This approach allows us to perform short integrations that end once the planets reach their current configuration.
\item Instead of relying solely on numerical integrations, we determine which dynamical processes affect the evolution of the KBOs and place constraints using analytical expressions. 
\item Instead of modeling all four planets directly, we model only Neptune but allow its orbit to change. We will demonstrate why this approach is sufficient for the problem we are exploring.
\end{itemize}

Our exploration of parameter space could produce two possible outcomes. If we find regions of parameter space that can deliver the hot objects on top of the cold, these consistent regions will provide constraints for more detailed models. If we rule out all of parameter space, then a new type of model, employing different physical processes, is necessary. Either way, we will identify and quantify what physical processes are responsible for sculpting the eccentricity distribution of the classicals in the generalized model we are treating. We emphasize that, rather than proposing a new model, we are exploring a generalization of Neptune's dynamical history, in which previous models correspond to a particular set of parameters.

In the next Section, we demonstrate that the hot and cold classicals have not only a bimodal inclination distribution, as already well established in the literature, but also distinct eccentricity distributions that were sculpted during Neptune's wild days. We use qualitative features of these eccentricity distributions to establish conservative criteria that models must meet. In Section 3, we establish the framework for our study and argue that, combined with the distinct physical properties of the hot and cold populations, these eccentricity distributions imply separate origins for the cold and the hot classicals. In  Section 4, we identify, for the classical region, the potential dynamical consequences of Neptune spending part of its dynamical history with high eccentricity --- delivery of objects via scattering, secular forcing, accelerated secular forcing near resonances, and a chaotic sea --- and present analytical expressions validated by numerical integrations. In Section 5, we combine the analytical expressions from Section 4 with the conservative criteria established in Section 2 to place constraints on Neptune's path, orbital evolution timescales, and interactions with other planets, ruling out almost all of parameter space. We find that Neptune must spend time with high eccentricity to deliver the hot classicals by these processes, but is restricted to one of two regions of $(a,e)$ space while its eccentricity is high. To avoid disrupting the cold objects, Neptune's eccentricity must have damped quickly or the planet's orbit must have precessed quickly while its eccentricity was high. Finally, because Neptune's current semimajor axis is ruled out when Neptune's eccentricity is high, Neptune is constrained to have migrated a short distance after its eccentricity damped. In the final Section, we discuss our results and their implications for the early history of the solar system.

\section{Constraints from the observed eccentricity distributions of hot and cold classicals}
\label{sec:obs}

\begin{figure}[htbp]
\begin{centering}
\includegraphics[width=.5\textwidth]{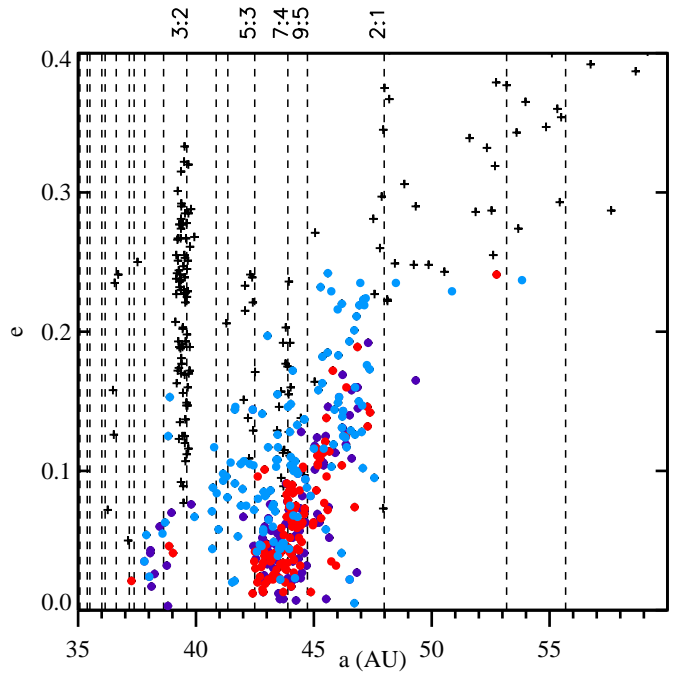}
\caption{Orbital eccentricity distribution of Kuiper Belt Objects. The resonant and scattered objects are plotted as black pluses. The classical objects are plotted as colored circles. The red objects have $i < 2^\circ$ and are thus very likely cold classicals. The blue objects have $i>6^\circ$ and are thus very likely hot classicals. The membership of any given purple object, which has $2^\circ<i<6^\circ$, is ambiguous (see Figure \ref{fig:cutoff}). Objects are taken from the Minor Planet Center Database and the Canada-France Ecliptic Plane Survey (CFEPS) and classified by \citet{2008G,2009K,2011V}. Dashed lines indicate the location of mean motion resonances with Neptune, which are included up through fourth order.}\label{fig:orbits}
\end{centering}
\end{figure}

The cold and hot populations are defined by the observed bimodal inclination distribution of classical KBOs \citep{2001B,2010G,2011V}. They also have distinct eccentricity distributions. The eccentricity distribution of all the observed KBOs is plotted in Figure \ref{fig:orbits}. In Section \ref{subsec:distinct}, we present evidence for distinct hot classical and cold classical eccentricity distributions and identify robust qualitative features of the distributions that models of Neptune's dynamical history must produce. In Section \ref{subsec:features} we lay out the observational constraints which we will use for the remainder of the paper.  In Section \ref{subsec:complications}, we assess the robustness of these features by performing statistical tests and considering observational bias.

\subsection{Evidence for distinct hot classical and cold classical eccentricity distributions}
\label{subsec:distinct}

\begin{figure}[htbp]
\begin{centering}
\includegraphics[width=.5\textwidth]{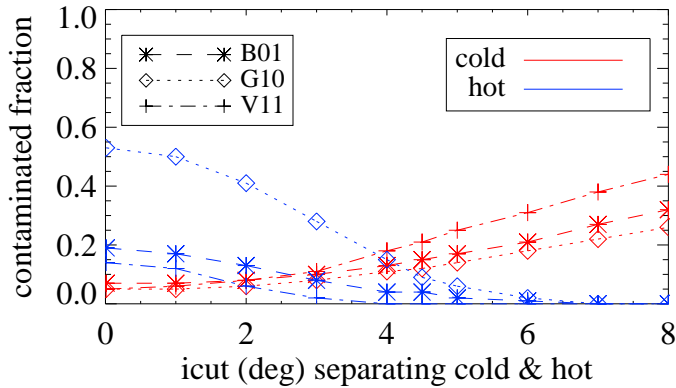}
\caption{Fraction of each population ``contaminated" by the other group as a function of the cut-off inclination $icut$ between the cold and hot population. The dashed lines, labeled B01, are calculated from the inclination distribution defined by \citet{2001B}; the dotted lines, labeled G10, from the inclination distribution defined by \citet{2010G}; and the dash-dotted lines, labeled V11, from the inclination distribution defined by \citet{2011V}.}\label{fig:cutoff}
\end{centering}
\end{figure}

We wish to use inclinations to separate the cold and hot classicals and then examine the eccentricity distributions of these two populations. Traditionally, the observed cold and hot objects have been separated using one inclination cutoff. However, because of the overlap between the hot and cold components in the bimodal inclination distribution, a single cutoff will necessarily result in the misclassification of hot objects as cold and vice versa. For example, if the classical population follows the model KBO inclination distribution derived by \citet{2010G} and we were to distinguish between the cold and hot populations using an inclination cut-off $icut = 4^\circ$, 11$\%$ of objects with $i<4^\circ$ would be actually be hot objects and 15$\%$ of objects with $i>4^\circ$ would be actually be cold objects. Thus, using $icut=4^\circ$, 11$\%$ of the objects classified as cold would be ``contaminated," and 15$\%$ of those classified as hot would be contaminated.  In Figure \ref{fig:cutoff}, we plot the ``contaminated" fraction over a range of values for $icut$ for the cold and hot populations based on three models of the debiased inclination distribution \citep{2001B,2010G,2011V}. For all three models, less than $10\%$ of the cold classicals are contaminated for $icut < 2^\circ$, while less $3\%$ of the hot classicals are contaminated for $icut > 6^\circ$. 

Therefore, instead of using a single $icut$, we divide the classicals into a likely cold population ($i<2^\circ$), a likely hot population ($i>6^\circ$), and an ambiguous population ($2^\circ<i<6^\circ$). We then examine the eccentricity distributions of the likely cold and likely hot populations, which are ``uncontaminated" samples. We use the uncontaminated eccentricity distributions to identify major features that models much match. In Appendix \ref{app:stats}, we confirm that our results are consistent if we probabilistically include the ambiguous population.

We wish to identify features of the eccentricity distribution that are sculpted during Neptune's wild days, \emph{not} by the long-term stability of the region under the influence of the modern solar system planetary configuration or by observational bias. First we compare the eccentricities of observed likely cold ($i < 2^\circ$) and likely hot ($i > 6^\circ$) objects to the survival map of \citet{2005L}, generated from a 4 Gyr simulation. \citet{2005L} generated initial conditions for test particles uniformly filling a cube of $(a, e, i)$ in the classical region: $41.375 AU < a <  48.125$ AU, $0 < e < 0.3$, and $0 < i < 30^\circ$. They then performed a 4 Gyr numerical integration including the test particles and the four giant planets (starting on their modern orbits). Then they computed the survival rate of test particles in bins of $(a, e)$ and $(a, i)$. In this work, we consider only the eccentricity survival map. This map bins over all inclinations\footnote{We note that at a given semimajor axis, the survival rate does not show a strong dependence on inclination \citep[][Figure 4, lower panel]{2005L}, except near the $\nu_8$ secular inclination resonance at 41.5 AU, which is devoid of low inclination objects. We do not establish constraints in this region.}. After 4 Gyr of evolution under the influence of the planets in their current configuration, KBOs with an initially uniform eccentricity distribution would be distributed according to this survival map. 

\begin{figure*}[htbp]
\includegraphics[width=\textwidth]{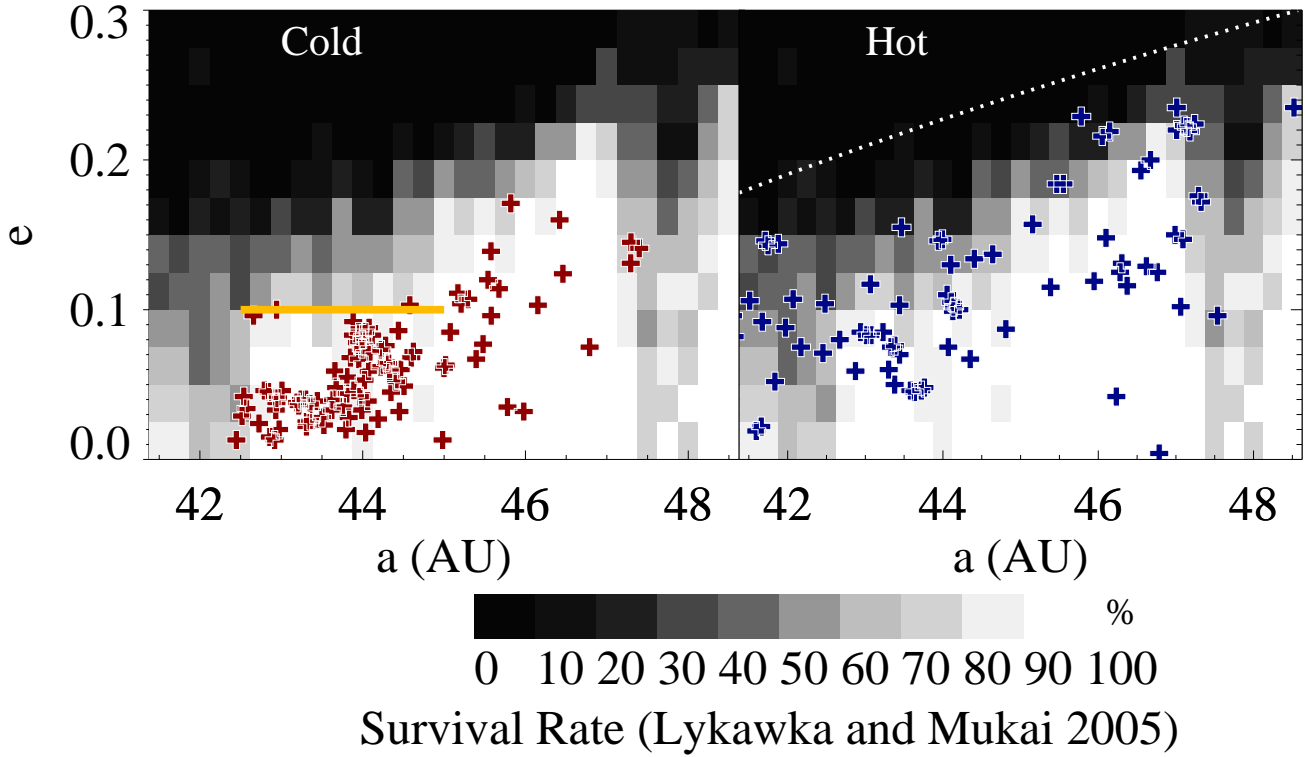}
\caption{Observed (plus symbol) eccentricity distributions of cold classicals with $i<^\circ2$ (left) and hot classicals with $i>6^\circ$ (right), plotted over the survival map of \citet{2005L}. In the left panel, the cold classicals between 42.5-44 AU are confined to $e < 0.05$, well below the survival limit in this region, while cold classicals between 44-45 AU are confined to $e < 0.1$, also below the survival limit. In the right panel, the hot classicals occupy the upper portion of the survival region. We plot $ e = 0.1$ from 42.5 to 45 AU as a solid yellow line in the left panel. The dashed line in the right panel, periapse $q = 34$ AU, indicates an approximate upper threshold of long-term survival, which we will use in Sections \ref{sec:hot} and \ref{sec:combo}.  Classical objects are taken from the Minor Planet Center Database and the Canada-France Ecliptic Plane Survey \citep[CFEPS;][]{2009K} and are classified by \citet{2008G}, \citet{2009K}, and \citet{2011V}.}\label{fig:diagnosticall}
\end{figure*}

However, rather than following the survival map, the observed cold and hot objects exhibit major distinct features. In Figure \ref{fig:diagnosticall}, we plot the sample of observed classical objects from the Minor Planet Center (MPC) and the Canada-France Ecliptic Plane Survey \citep[CFEPS;][]{2009K} on top of the \citet{2005L} stability map. The cold objects are confined to very low eccentricities. From 42.5 to 44 AU, the cold objects appear to be confined to $e < 0.05$. From 44 to 45 AU, the cold objects appear confined below $e < 0.1$. This confinement of the cold classicals to below the survival limit implies that they were not excited above these levels because if they had been, we would still observe objects at higher eccentricities. Similarly, \citet{2009K} found that classical objects with $i<4.5^\circ$ are restricted to $42.5$ AU $< a < 45$ AU and $e < 0.1$. In contrast, hot objects occupy the upper portion of the survival region and appear uniformly distributed in $a$ from 42 to 47.5 AU. Suggestively, they also appear to be distributed roughly along a scattering line, as if they were scattered into the classical region but did not have time to evolve to low eccentricities before Neptune's eccentricity damped.

\subsection{Conservative criteria that models must meet}
\label{subsec:features}

We use the following major qualitative features to place constraints on Neptune's dynamical history. We consider these criteria ``conservative" because they allow for dynamical histories at the very edge of consistency with the observations.

\emph{Cold population:  confined to low eccentricities of $e < 0.1$ in the region from 42.5 to 45 AU.}  In the region between 42.5 and 45 AU, the cold objects have eccentricities well below the distribution that follows the survival map. Therefore, Neptune cannot excite the cold classical objects in this region above $e = 0.1$. (We choose this value to be conservative in ruling out regions of parameter space and to match \citet{2009K}, but it appears that cold objects with semimajor axes less than 44 AU are confined below $e < 0.05$, a tighter constraint.) We indicate this threshold as a solid yellow line in Figure \ref{fig:diagnosticall}.

\emph{Hot population: delivered to the upper survival region with $q > 34$ AU out to 47.5 AU.} The observed hot objects occupy the upper portion of the survival region. Therefore, a consistent dynamical history should allow some objects to reach this region. It is not necessary for the transported objects to reach very low eccentricities, only low enough to survive under the current planetary configuration. We set the criterion that the hot classicals must be delivered to periapse $q > 34$ AU (dashed line in Figure \ref{fig:diagnosticall}) from 42 A to 47.5 AU, the edge of observed population. 

\subsection{Assessing the robustness of the observed features}
\label{subsec:complications}
In determining which major features serve as constraints on the dynamical history of the solar system, we address several complications:

\begin{enumerate}
\item The inclinations of objects vary over time \citep{2011V}.
\item The inclination cut-off between the hot and cold classicals is model dependent.
\item Proper elements are more robust than the observed instantaneous elements.
\item The features in the eccentricity distributions might be the result of random chance or small number statistics.
\item The eccentricity distributions may be impacted by observational bias.
\end{enumerate}

The first complication is addressed by \citet{2011V}. They find that, at any given time, only $5\%$ of objects will be inconsistent with their original inclination-based classification of hot versus cold. Therefore we expect the major qualitative features we identify to hold despite variations in the inclinations of some objects

The second complication is that the cut-off inclination between the hot and cold classicals depends on the parameters and form of the model for the bimodal inclination distribution. The three models we consider \citep{2001B,2010G,2011V} each use the functional form of $\sin i$ multiplied by a Gaussian but use different widths and cold/hot fractions. They also use different planes for the inclination: \citet{2001B} defines the inclination with respect to the ecliptic plane, \citet{2010G} with respect to the mean plane of the Kuiper Belt, and \citet{2011V} with respect to the invariable plane. However, despite these differences, $i < 2^\circ$ and $i > ^\circ6$ are robust cut-offs for establishing an uncontaminated cold and hot population, respectively, for each of the three models. In regards to the functional form of the model, \citet{2011V} find that the high-inclination component is not well-described by a Gaussian, and \citet{2009F} argue that the most robust generic functional form for a distribution of inclinations is a Fischer distribution. However, the discrepancies between different functional forms are strongest for classifying objects in the intermediate, overlapping portion of the bimodal inclination distribution (K. Volk 2011, private communication), so we argue that our approach of definitively classifying only the ``uncontaminated" low and high inclination objects is robust.

Regarding the third complication, the stability map of \citet{2005L} is formulated in terms of instantaneous eccentricity and inclination, but the most robust, non-varying formulation of the orbital elements are the proper, or free, elements. However, none of the model inclination distributions are formulated in terms of the proper inclination, nor is the stability map of \citet{2005L} formulated in the proper elements. To compare ``apples to apples," we use the instantaneous orbital elements in the plots in this section. We use the proper elements in Appendix \ref{app:stats} and find that the observational features we identify (Section \ref{subsec:features}) still hold.

Addressing the fourth complication, in Appendix \ref{app:stats}, we confirm that the confinement of the cold population to low eccentricities is statistically significant. For the hot population, we only impose the constraint that the models must deliver them to the long-term stable region (Section \ref{subsec:features}); we will demonstrate that this loosely formulated restriction ends up imposing strong constraints on Neptune's dynamical history.

\subsubsection{Ruling out observational bias through statistical tests}
\begin{figure*}[htbp]
\includegraphics[width=\textwidth]{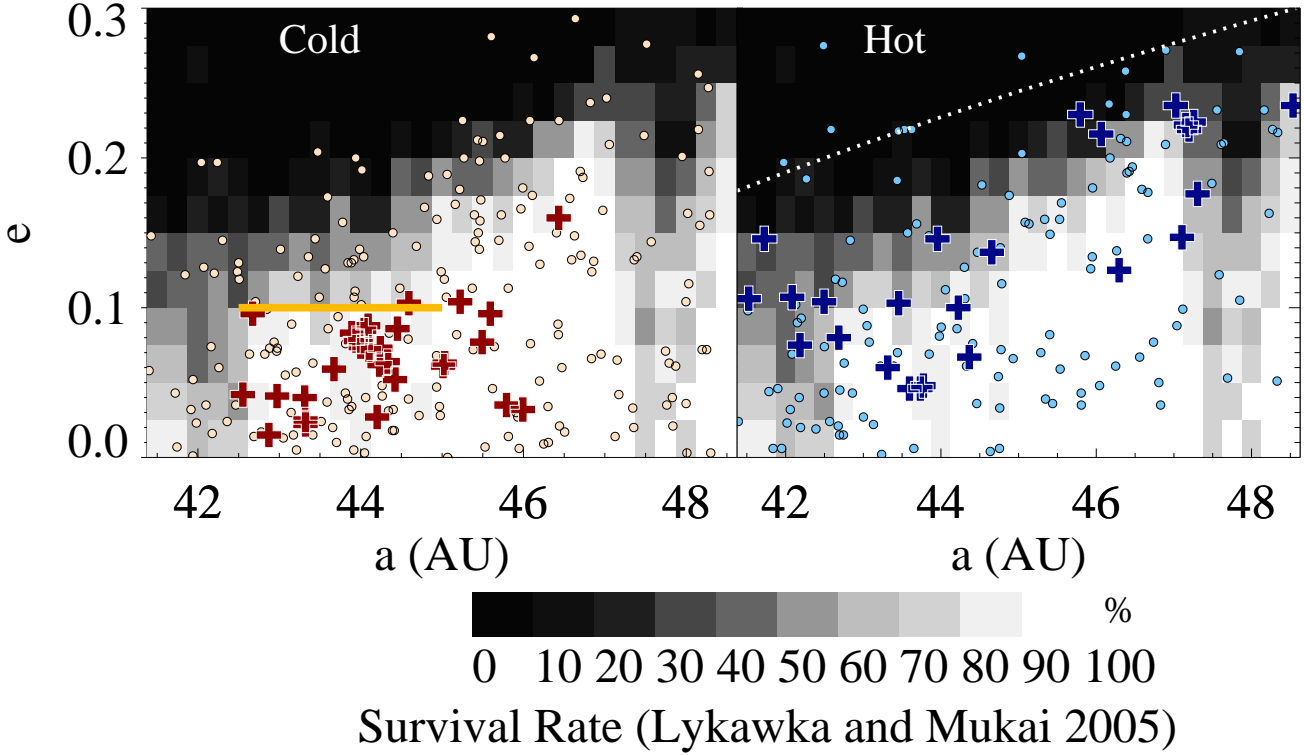}
\caption{CFEPS objects only. Plotted over the survival map of \citet{2005L} are predicted (circle) and observed (plus) distributions of cold classicals with $i<2^\circ$ (left) and hot classicals with $i>6^\circ$ (right). The predicted classicals are the distribution expected from a uniform $(a,e)$ distribution, filtered by the survival map and put through the CFEPS Survey Simulator of \citet{2009K} (see the text for further detail). The observed classicals are those observed by CFEPS. In the left panel, the cold classicals between 42.5 and 44 AU are confined to $e < 0.05$, well below the survival limit, while cold classicals between 44 and 45 AU are confined to $e < 0.1$, also below the survival limit in this region. We plot $ e = 0.1$ from 42.5 to 45 AU as a solid yellow line in the left panel. In the right panel, the hot classicals occupy the upper portion of the survival region. The dashed line indicates an approximate upper threshold of long-term survival, which we will use in Sections \ref{sec:hot} and \ref{sec:combo}.}\label{fig:diagnostic}
\end{figure*}

We would not expect observational bias (the final complication) to cause the cold classicals to appear to confined to low eccentricities; KBO searches are more likely to preferentially observe high-eccentricity (i.e. small periapse) objects of a given semimajor axis. However, to ensure that the features on which we base our constraints (Section \ref{subsec:features}) are not created by observational bias, we perform the following test to see if observational bias could generate them:
\begin{enumerate}
\item We begin by generating a simulated sample of objects uniformly distributed in $(a, e)$. We set the inclinations to follow the unbiased inclination distribution of the classicals, as modeled by \citet{2010G}.
\item Then we use the stability map of \citet{2005L} to transform this simulated sample following a uniform eccentricity distribution into a sample following the eccentricity distribution shaped by the four giant planets under their current configuration. We call this process ``filtering." For each simulated object, we obtain a predicted survival rate from the $(a,e)$ stability map of \citet{2005L}. Then we select a uniform random number between 0 and 1. If the randomly selected number is less than the predicted survival rate, we include the object in our sample. Note that since the survival rates of \citet{2005L} are given as a range (e.g. $10-20\%$, $90-100\%$), we repeated this entire test (i.e., steps 2-4) three times, once using the minimum of each range, once using the mean of each range, and once using the maximum of each range. As expected, the resulting eccentricity distribution had higher (lower) eccentricities when we used the maximum (minimum) each range but the major features we identified still held. The simulated population in Figure \ref{fig:diagnostic} uses the mean.
Next we transformed the ``survival-rate filtered" sample from step 2 into an observed sample:
\item We randomly assign each object an \emph{H} magnitude\footnote{An alternative method, randomly drawing the \emph{H} magnitudes from observed classical CFEPS objects, yielded results that were qualitatively the same.} between 6 and 8. 
\item Then we apply the L7 Survey Simulator for the well-characterized CFEPS.  \citep{2009K}. 
\item We compare the final simulated distribution to the subset of objects that were detected by CFEPS \citep{2009K} (Figure \ref{fig:diagnostic}). Figure \ref{fig:diagnostic} is analogous to Figure \ref{fig:diagnosticall}. It includes the simulated distribution (circles) and only the subset of KBOs observed by CFEPS. The simulated distribution is not confined to $e < 0.1$ from 42.5 to 45 AU, confirming this feature of the observed eccentricity distribution does not result from observational bias. Note also that simulated hot objects are found at lower eccentricities than observed.
\end{enumerate}

\section{Framework}
\label{subsec:single}

Our representation of the observations in Figure \ref{fig:diagnosticall} highlights the problem with theories of a single origin for the hot and cold objects. How could a single origin produce both a cold population confined to low eccentricities and a hot population, with different physical properties and inclinations, dwelling above at high eccentricities? In Section \ref{subsubsec:single}, we explain why a single origin scenario is unlikely. In Section \ref{subsec:double}, we describe a scenario, which we will consider throughout the rest of the paper, in which the cold population formed in situ and the hot population  formed in the inner disk and was transported to the classical region. In Section \ref{subsec:short}, we explain why is it reasonable to place constraints on Neptune's history using the evolution of the KBOs only during Neptune's wild days, and in Section \ref{subsec:alternative} we discuss the possibility of alternative scenarios of Kuiper Belt assembly.

\subsection{Ruling out a single origin for the hot and cold classicals}
\label{subsubsec:single}
A single origin for the hot and cold populations seems unlikely. If the cold and hot classicals formed together in the classical region, where they are observed today, it is difficult to imagine a process that would excite the hot population while leaving the cold population confined to low eccentricities. \citet{2005H} proposed a scenario in which the classical region has been pre-excited. However, this scenario does not produce a population of cold classicals confined to low eccentricities. Moreover, if both the cold and hot classicals were transported from the inner disk, it seems unlikely that a common deposition process would place the cold classicals solely at low eccentricities. \citet{2003L} and \citet{2008L} propose scenarios in which both the hot and cold classicals are transported from the inner disk. 

In the scenario of \citet{2003L}, the cold classicals were pushed outward by the 2:1 resonance and dropped during stochastic migration, while the hot classicals scattered off of Neptune. The feasibility of this mechanism depends on the size distribution of planetesimals, because the migration needs to be stochastic in order to drop objects from resonance. When Neptune scatters a planetesimal inward and the planetesimal is ejected by Jupiter, Neptune experiences a net gain in angular momentum and migrates outward. If the planetesimals are small, this is a smooth process, but if they are large, it is a jumpy, stochastic process, in which KBOs can be dropped from resonance. See \citet{2006M} for a detailed exploration of stochastic migration; they conclude that planetesimal-driven migration cannot generate the necessary stochasticity unless a large fraction of planetesimals formed very large. This constraint merits a fresh look in light of new planetesimal formation models \citep[see][and references therein]{2010C}. However, even if extreme planetesimal properties allowed this mechanism to work, objects dropped from the 2:1 resonance would have a range eccentricities, not be confined solely to low eccentricities. Therefore this mechanism holds more potential for producing the hot population than the cold population.

In the scenario of \citet{2008L}, the cold classicals are objects that, like the hot classicals, were scattered into the Kuiper Belt by an eccentric Neptune but, unlike the hot classicals, evolve down to low eccentricities in regions near resonances. However, this mechanism would create a range of eccentricities for the cold classicals and thus would have trouble producing the confined eccentricities of the region of $42.5$ AU $< a < 45$ AU (Figure \ref{fig:diagnosticall}). They do find some correlation between a particle's final inclination and its initial semimajor axis in one of their simulations \citep[][Figure 11, panel (b)]{2008L}, which may be able to partially account for a difference in physical properties between low and high inclination objects. However, transporting the cold classicals from the inner disk  is not consistent with the finding by \citet{2010P} that wide binaries --- of which the cold population contains a number --- cannot survive transportation from the inner disk to the classical region.

\subsubsection{The eccentricity distribution was not sculpted solely by a different stability threshold in the past}
\label{subsec:notstability}

One might wonder whether the observed confinement of the cold classicals (Figure \ref{fig:diagnosticall}) is the result of a smaller stability region than exists today, as if the cold classicals follow an ancient scattering line. However, there are numerous ``hot" objects with $i > 6^\circ$, as well as ambiguous objects with $2^\circ < i < 6^\circ$, in the region from 42.5 to 45 AU that have high eccentricities, right up to the modern stability limit. To create the observed distribution, one would need a mechanism that removes all objects with high eccentricities and $i < 2^\circ$ while leaving a) objects with low eccentricities and $i < 2^\circ$, and b) objects with a range of eccentricities and $i > 2^\circ$. Therefore, it seems unlikely that this mechanism could produce both the hot and cold populations. We note that such a scenario could take place before Neptune transports the hot classicals. However, such initial sculpting would not affect the constraints we will place, which Neptune still needs to obey during the hot classical transport phase.

\subsection{Colds in situ, hots transported from the inner disk}
\label{subsec:double}
\begin{figure*}[htbp]
\begin{centering}
\includegraphics[width=\textwidth]{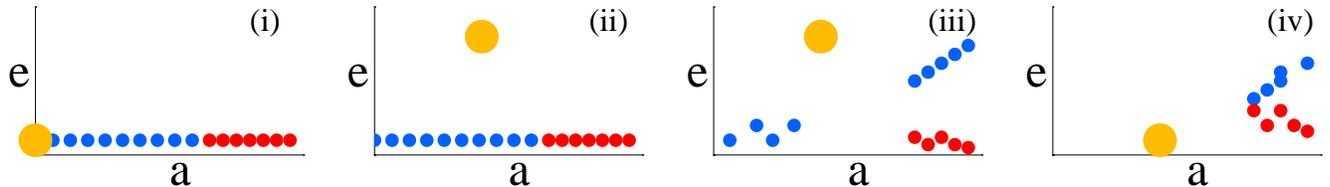}
\caption{Conceptual framework. (i) The hot classicals (blue) form in the inner disk and the cold classicals (red) form in the outer disk. (ii) Neptune is scattered onto a high-eccentricity orbit through its interactions with other planets. (iii) Neptune scatters the hot objects into the classical region without disrupting the cold ones. (iv) Neptune's large eccentricity damps, leaving the cold classicals confined to low eccentricities.}\label{fig:concept}
\end{centering}
\end{figure*}

Thus, throughout the rest of paper, we consider the general scenario --- also discussed in \citet{2008M} --- in which the hot objects are transported to the classical region from the inner disk and the cold objects form in situ in the classical region. The cold objects must not be dynamically excited, as quantified by the criterion we established in Section \ref{subsec:features}. In Figure \ref{fig:concept}, we show a conceptualization of this model. This general scenario encompasses previous models and allows Neptune to undergo any potential combination of high eccentricity, migration, and/or apsidal precession with a range of initial eccentricities and semimajor axes. If our constraints do not rule out all of parameter space for this model, it may be possible to produce both the hot and cold classical population. Otherwise, a major physical process is missing from current models. 

For now, we can think of Neptune as having a high eccentricity at one location after undergoing planet-planet scattering. Our results hold in more complicated scenarios as well, as we will describe in Section \ref{subsec:combointerpret}.

\subsection{The case for considering short-term evolution}
\label{subsec:short}

We have chosen features (Section \ref{subsec:features}) that are not shaped by the long-term survival of KBOs under the current solar system planetary configuration. Therefore, we can focus on modeling the processes that affect these features during the interval of Neptune's wild days instead of treating the entire 4 Gyr. We model these processes analytically in Section \ref{sec:process} and validate our analytical expressions using numerical integrations.

Unless otherwise specified, the integrations are performed as follows. We perform the integrations using the \emph{Mercury 6.2} hybrid symplectic integrator \citep{1999C}, an \emph{N}-body code that allows massless test particles. We employ an accuracy parameter of $10^{-12}$ and a step size of 200 days and impose (if applicable) the migration and damping of Neptune's orbit through user-defined forces and velocities. Migration and damping follow the following forms:
\begin{eqnarray}
\label{eqn:forms}
\aN = (\aN)_f + ((\aN)_{0} - (\aN)_{f} ) \exp{(-t/\tau_{\aN})} \nonumber\\
\eN = (\eN)_0 \exp{(-t/\tau_{\eN})} \nonumber,\\
\end{eqnarray}
\noindent where $\aN$ is the semimajor axis of Neptune at time $t$, $(\aN)_0$ is the initial semimajor axis, and $(\aN)_f$ = 30.1 AU is the final semimajor axis. At time $t$, the eccentricity of Neptune is $\eN$; Neptune's initial eccentricity is $(\eN)_0$. The forced evolution of Neptune's orbit is implemented through modifications to \emph{Mercury 6.2}, described in detail in the Appendix of \citet{2012W}. The migration and damping are parametrized by timescales $\tau_{\aN}$ and $\tau_{\eN}$ respectively, which we specify in the text in the applicable cases. When specified, Neptune is forced to undergo apsidal precession using an artificial stellar oblateness force, built into \emph{Mercury 6.2}, that we modified to apply only to Neptune, parameterized by a $J_2$ coefficient chosen to produce the correct precession rate. Unless otherwise noted, Neptune is the only planet included in the integration. The KBOs are modeled as 600 massless test particles with initial $a$ evenly spaced between 40 and 60 AU and initial $e = i = 0$. The migration, damping, and apsidal precession are not applied to the KBOs, only to Neptune.

\subsection{Alternative scenarios for Kuiper Belt assembly}
\label{subsec:alternative}

The purpose of this paper is to explore whether the generalized scenario described here can even work, i.e. whether it is ever possible to transport the hot classicals from the inner disk to the classical region without disrupting an in situ cold population. Yet alternative scenarios exist that do not fit within this framework, such as the additional planet beyond Pluto proposed by \citet{2008LM} and others. Furthermore, we will describe in the conclusion how additional constraints could rule out the generalized model we consider. In that case, development of alternative scenarios would be necessary. Obviously, the constraints we will place do not necessary hold for a scenario that is not encompassed by our general model.

\section{Physical processes resulting from Neptune's high eccentricity}
\label{sec:process}
We begin our analysis by describing the physical processes that can impact the Kuiper Belt if Neptune's eccentricity is high. First we consider how classical KBOs reach the classical region. In our generalized model (Figure \ref{fig:concept}), the cold classicals form in situ and hot classicals are delivered by Neptune from the inner disk. Once the hot and cold objects are in the classical region, both evolve in response to an eccentric Neptune. A tension arises between need of hot objects for Neptune to be eccentric --- to deliver them into the classical region and to allow them to evolve to low eccentricities once they arrive --- and the undesirable excitation of cold objects when Neptune is eccentric. In this Section, we lay out analytical expressions for how KBOs evolve in response to an eccentric Neptune and use this theory to transform our observational constraints into comprehensive constraints on Neptune's orbit during its high eccentricity period. We will employ these constraints on Neptune's orbit in Section \ref{sec:constraints} to rule out much of parameter space.

\subsection{Delivery into the classical region}
\label{subsec:scatter}

In the generalized model we explore, Neptune may be scattered outward from the inner solar system onto a highly eccentric orbit. After this occurs, Neptune's new orbit crosses the orbits of some planetesimals in the inner disk (see Figure \ref{fig:concept}), which scatter off the planet. This mechanism can potentially deliver hot objects from the inner disk into the classical region.

\begin{figure}[htbp]
\begin{centering}
\includegraphics[width=.5\textwidth]{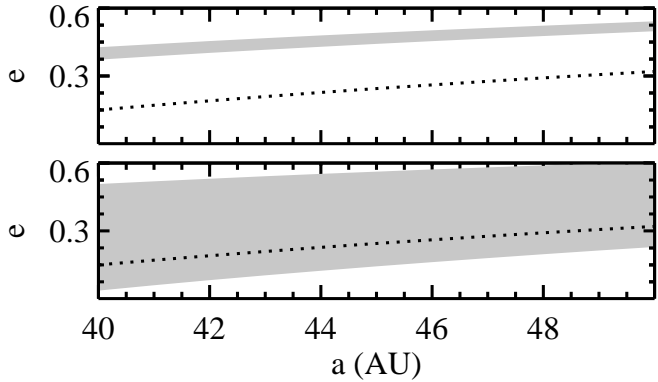}
\caption{Region into which Neptune can scatter particles for $\aN=24, \eN = 0.02$ (top), a typical initial condition for extensive migration models, and $\aN=28.9, \eN = 0.3$ (bottom), which is the initial condition for Run B in \citet{2008L}. The dashed line marks the upper threshold of long-term survival, as indicated in Figure \ref{fig:diagnostic}. } \label{fig:scatterallow}
\end{centering}
\end{figure}

The region into which Neptune can scatter objects is defined by the planet's semimajor axis $\aN$ and eccentricity $\eN$. Neptune can scatter objects outward to periapses $q = a(1-e)$ between Neptune's periapse $r_{p,{\rm N}}= (\aN - a_{\rm H}) (1-\eN)$ and apoapse $r_{a,{\rm N}} =(\aN + a_{\rm H})  (1+\eN)$. In Figure \ref{fig:scatterallow}, we show examples of the region into which Neptune can scatter KBOs for two sets of parameters $(\aN, \eN)$. We have adjusted $r_{p,{\rm N}}$ and $r_{a,{\rm N}}$ to include the Hill sphere radius, $a_{\rm H}$ ($\sim$ 1 AU), the distance from Neptune at which Neptune's gravity overcomes the Sun's tidal gravity. Particles that enter Neptune's Hill sphere will be scattered and, if they are scattered outward into the classical region, will reach -- under the approximation that the scattering location becomes the particle's new periapse -- a given semimajor axis $a$ with eccentricities between $1-r_{p,{\rm N}}/a$ and $1-r_{a,{\rm N}}/a$.

\subsubsection{Implications of scattering for the cold classicals}
\label{subsec:coldscatter}
If Neptune's apoapse $r_{a,{\rm N}}$ is large enough, the planet can potentially impinge into the cold classical region, scattering the objects there. However, we find in practice that if Neptune's eccentricity damps within the constraints we will place, then cold objects are never excited up to the scattering line, which moves upward as Neptune's eccentricity damps. In Section \ref{subsec:precess}, we will return to this issue for the special case of Neptune undergoing fast apsidal precession.

\subsection{Secular forcing}
\label{sec:theory}

Secular forcing has a strong effect on the behavior of KBOs during the potential period when Neptune's eccentricity is high. (See \citealt{2000M} for a pedagogical description of first order secular theory outside of mean motion resonance.) On a timescale of a million years, a cold object is excited to a higher eccentricity via Neptune's secular forcing. (The direct forcing from the other planets on the KBO is negligible and the other planets only affect the KBOs via Neptune, as we will demonstrate in Section \ref{subsec:otherplanets}). A hot object that is scattered into the classical region also experiences secular forcing, which can decrease its eccentricity so that its orbit no longer crosses Neptune's. Hot or cold, an object's eccentricity is a vector combination of its forced eccentricity --- set by Neptune's eccentricity, Neptune's semimajor axis relative to the object's, and Neptune's apsidal precession rate --- and of the object's free eccentricity, which is set by its initial condition before Neptune is scattered to a high eccentricity. The object's free eccentricity precesses about the forced eccentricity at the secular frequency $\gkbo$. Therefore, as we will demonstrate, a cold object has a well-defined excitation time and amplitude, and a hot object will have a minimum eccentricity it can reach after being scattered into the classical region. Thus while Neptune's eccentricity is high, secular forcing potentially is an important mechanism for exciting the cold objects and stabilizing the hot ones. As we will show, when Neptune's eccentricity damps quickly, the orbits of the KBOs are ``frozen" near the eccentricities they reached through secular evolution.

\subsubsection{Basic secular evolution}
\label{subsec:basic}

First we define expressions for the secular evolution of a test particle under the influence of a planetary system containing only Neptune and the Sun. The components of the particle's eccentricity vector are $h = e \sin \varpi$ and $k = e \cos \varpi$, where $\varpi$ is the particle's longitude of periapse. Secular forcing by Neptune causes $h$ and $k$ to evolve as (to first order in $e$ and $\eN$):
\begin{eqnarray}
\label{eqn:sec}
h = e_{\rm free} \sin(\gkbo t + \beta) + e_{\rm forced}  \sin (\varpiN) \nonumber \\
k = e_{\rm free} \cos(\gkbo t + \beta) + e_{\rm forced} \cos (\varpiN) \nonumber \\
\end{eqnarray}
where 
\begin{eqnarray}
\label{eqn:simples}
e_{\rm forced} =   \frac{b_{3/2}^{(2)}(\alpha)}{  b_{3/2}^{(1)}(\alpha) } \eN \nonumber \\
\alpha_= \frac{\aN}{a} \nonumber \\
\gkbo = \alpha b_{3/2}^{(1)}(\alpha)  \frac{\mN}{m_\bigodot}   \frac{n}{4}   \nonumber \\
\end{eqnarray}

The constants  $e_{\rm free}$ and $\beta$ are determined from the initial conditions, and the particle's forced eccentricity is $e_{\rm forced}$. Here, $\varpiN$ is the longitude of periapse of Neptune, $\eN$ is the eccentricity of Neptune, and $\alpha$ is the ratio of Neptune's semimajor axis to that of the particle, all of which are assumed to be constant. The functions $b$ are standard Laplace coefficients (see \citealt{2000M}). The secular frequency of the KBO is $\gkbo$, $\mN$ is the mass of Neptune, $m_\bigodot$ is the mass of the Sun, $n = (Gm_\bigodot/a^3)^{1/2}$ is the particle's mean motion, and $G$ is the universal gravitational constant.

Consider a cold object with $e=0$ at $t=0$. In our approximation, Neptune, having been scattered by the other giant planets, effectively instantaneously appears and imparts a forced eccentricity of $e_{\rm forced}$. From these initial conditions, $e_{\rm free} = e_{\rm forced}$. Then, the KBO's forced eccentricity vector remains fixed and the object's total free eccentricity vector precesses about the forced eccentricity. Thus its total eccentricity varies sinusoidally from $e=0$ to $e = 2 e_{\rm forced}$ on a timescale set by $\gkbo$. 

A hot object scattered into the classical region, in contrast,  has an eccentricity $e$ at $t=0$. The magnitude of its free eccentricity is a value between max(0, $e - e_{\rm forced}$) and $e + e_{\rm forced}$, depending on the initial location of its periapse relative to Neptune's. Over a timescale set by $\gkbo$, its total eccentricity oscillates. Depending on the initial conditions, it may reach an eccentricity low enough so that its orbit no longer crosses Neptune's and/or so that it is stable under the current configuration of the giant planets.

\begin{figure*}[htbp]
\begin{centering}
\includegraphics[width=\textwidth]{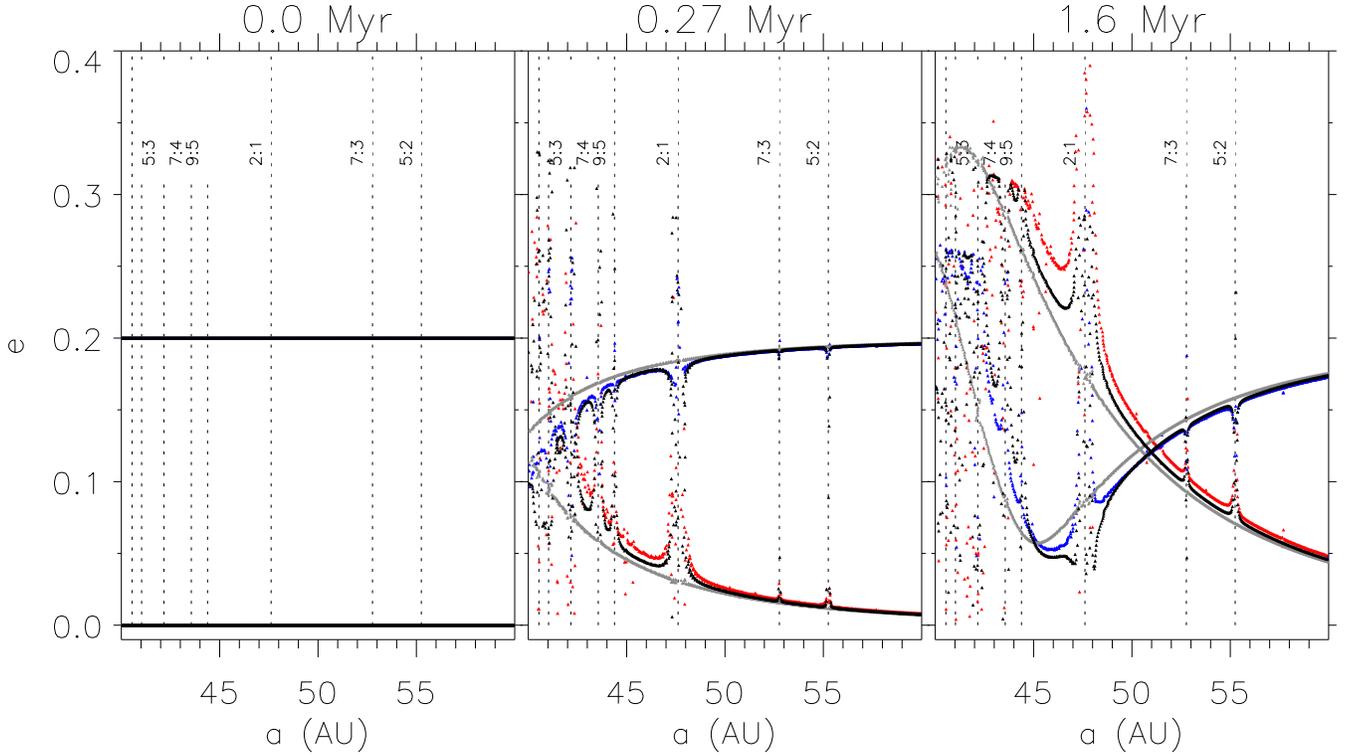}
\caption{Numerical integration ($\aN = 30, \eN = 0.2$) shows cold objects (red) secularly evolving to high eccentricities and hot objects (blue) to low eccentricities. The gray is our analytical model without the resonance correction terms (Equation \ref{eqn:sec}) and the black, which matches much better near the resonances, includes resonance terms (Equation \ref{eqn:effects}). \label{fig:sec}}
\end{centering}
\end{figure*}

An example of the secular evolution of cold objects ``going up" and hot objects ``going down" in eccentricity is shown in Figure \ref{fig:sec}, highlighting the tension between the evolution of hot objects to low eccentricities and the evolution of cold objects to high eccentricities. The cold objects (red) begin with $e = 0$ (see Section \ref{subsec:short} for a general description of the integrations we performed.) The hot objects in the integration (blue) all begin with $e = 0.2$ and $\varpi = \varpi_{\rm N} + \pi/3$, for the purposes of illustrating secular evolution\footnote{As shown in Section \ref{subsec:scatter}, a real hot object can only be scattered to a certain region of $(a,e)$ space in the classical region, and its eccentricity and periapse are actually correlated}. As time progresses through three snapshots, the cold objects become excited and the hot objects reach low eccentricities. The analytical model from Equation (\ref{eqn:sec}) matches well except near mean motion resonances, where the secular evolution is much faster than predicted. We also overplot a more accurate analytical expression that includes a resonant correction term, which we will derive in Section \ref{subsec:refined}.

We note that throughout the paper, we perform numerical integrations of objects with initial semimajor axes out to 60 AU to give a better conceptual picture of the secular excitation. Moreover, depending on where the initial population was truncated in the solar system's planetesimal disk, it is possible that additional classical KBOs will be discovered beyond 48 AU in the future, and we would like to make testable predictions. Finally, the integration results should be interpreted as examples: since we cannot show a figure for every possible combination of parameters for Neptune, we plot out to 60 AU to let the reader imagine the results if Neptune's semimajor axis were smaller.

\subsubsection{Refined secular expression}
\label{subsec:refined}

The secular expression in Section \ref{subsec:basic} is valid to first order in eccentricity, neglects the effects of orbital resonances, and applies in the case in which Neptune's orbit does not apsidally precess. However, these neglected effects can significantly alter a secularly evolving KBO's behavior:

\begin{enumerate}
\item Neptune's high eccentricity makes terms of order $\eN^2$ (ignored in deriving Equation \ref{eqn:sec}) non-negligible. As we will show, these extra terms result in a faster secular forcing frequency of the KBO.
\item Proximity to mean-motion resonances with Neptune significantly alters the secular frequencies of KBOs \citep[as shown for other solar system bodies in][]{1897H,1989M,2011M}. Following \citet{1989M}, we incorporate resonance correction terms, described in our Appendix \ref{subsec:resterms}. These resonance correction terms are very important for objects \emph{near} resonance but not valid for objects librating in resonance (we are not considering resonant objects in this paper\footnote{It has been claimed \citep[e.g.][]{2008L} that the entire classical region from the 3:2 to the 2:1 resonance is full over overlapping resonances. However, in Section \ref{subsec:chaos} we will demonstrate that Neptune's high eccentricity alone does not cause resonance overlap in the classical region.}).
\item As Neptune's eccentricity is damped, the particle's forced eccentricity goes to zero. If the damping occurs over a timescale $\tau_{\eN}$ shorter than the secular oscillation time, the particle's eccentricity is frozen at approximately the value it reaches at the eccentricity damping time. If the damping occurs over a longer timescale, the particle's eccentricity evolves to its initial free eccentricity.
\item Apsidal precession of Neptune alters the forced eccentricity, keeping it low when Neptune precesses quickly. It also alters the oscillation timescale of the total eccentricity, because now both the free and forced eccentricity are precessing.
\item Migration alters the secular frequencies and shifts the locations of the resonances.
\end{enumerate}

Based on these considerations, we modify the standard expression (Equation \ref{eqn:sec}) for the secular evolution of a test particle's $h$ and $k$ under the influence of an eccentric Neptune to incorporate these effects. 
\begin{eqnarray}
\label{eqn:effects}
h = e_{\rm free, 0} \sin(\gkbo t + \beta_0) + \bar{e}_{\rm forced}  \sin (\varpi_{{\rm N},0} + \dotvarpiN t) \nonumber \\
k = e_{\rm free, 0} \cos(\gkbo t + \beta_0) + \bar{e}_{\rm forced} \cos (\varpi_{{\rm N},0}+ \dotvarpiN t) \nonumber \\
\end{eqnarray}
\begin{eqnarray}
\gkbo = (1+\frac{f_5}{f_2}\eN^2)\alpha b_{3/2}^{(1)}(\alpha) \frac{\mN}{m_\bigodot}   \frac{n}{4} + \epsilon \delta \gkbo \label{eqn:extra} \\
\delta \gkbo = \alpha (C x \eN^{x-1})^2  \frac{\mN}{m_\bigodot} n \label{eqn:extradelta} \\
\bar{e}_{\rm forced} =  \sin ({\rm min} (g_{\gkbo} \tau_{\eN},\pi/2))  \frac{\gkbo'}{\dotvarpiN - \gkbo  } \eN(t) \label{eqn:extraforced}\\
\gkbo' =   -  (1+\frac{f_{10}}{f_{11}}\eN^2)\ \alpha b_{3/2}^{(2)}(\alpha) \frac{\mN}{m_\bigodot}   \frac{n}{4} \label{eqn:extrafactor}
\end{eqnarray}

Compare Equation (\ref{eqn:effects}) to Equation (\ref{eqn:sec}). The form is the same but Equation (\ref{eqn:effects}) has several key differences and new variables, that we will proceed to discuss and define throughout the remainder of this subsection. One distinction is that several quantities that were fixed in Equation (\ref{eqn:sec}) (i.e., $\varpiN, \alpha, \eN$) can now vary with time. Additionally, we have incorporated corrections for Neptune's high eccentricity and for the potential proximity of the KBO to mean motion resonance with Neptune. The impact of the time-varying $\eN$ is the most complicated, so we leave it for last.

As our first correction, we allow the longitude of pericenter of Neptune, $\varpiN$, to precess. Neptune's precession adds an extra term  $\dotvarpiN$ to Equation (\ref{eqn:effects}) (where we rewrite $\varpiN$ as a linear function of time: $\varpi_{{\rm N},0} + \dotvarpiN t$) and to Equation (\ref{eqn:extraforced}), in analogy to the standard secular theory for the four-planet case. We note that the other giant planets impact the secular evolution of the KBOs indirectly by causing Neptune's eccentricity to precess (see Section \ref{subsec:otherplanets} for discussion). 

The precession rate $\dotvarpiN$ has the same role (and same place, in the denominator of the forced eccentricity) in the just-Neptune secular theory as in the standard four-planet secular theory (Appendix \ref{app:planets}) for the particle except that we are specifying Neptune's evolution via $\dot{\omega_{\rm N}}$ instead of constructing a secular theory for the planets that produces a particular $\dot{\omega_{\rm N}}$. When $\dotvarpiN$ is large, the effective forced eccentricity $e_{\rm forced} = |\bar{e}_{\rm forced} |$, defined below, remains low because the forced eccentricity is inversely proportional to the precession rate for $|\dotvarpiN | \gg | \gkbo|$ (Equation \ref{eqn:extraforced}).

\begin{figure*}[htbp]
\begin{centering}
\includegraphics[width=\textwidth]{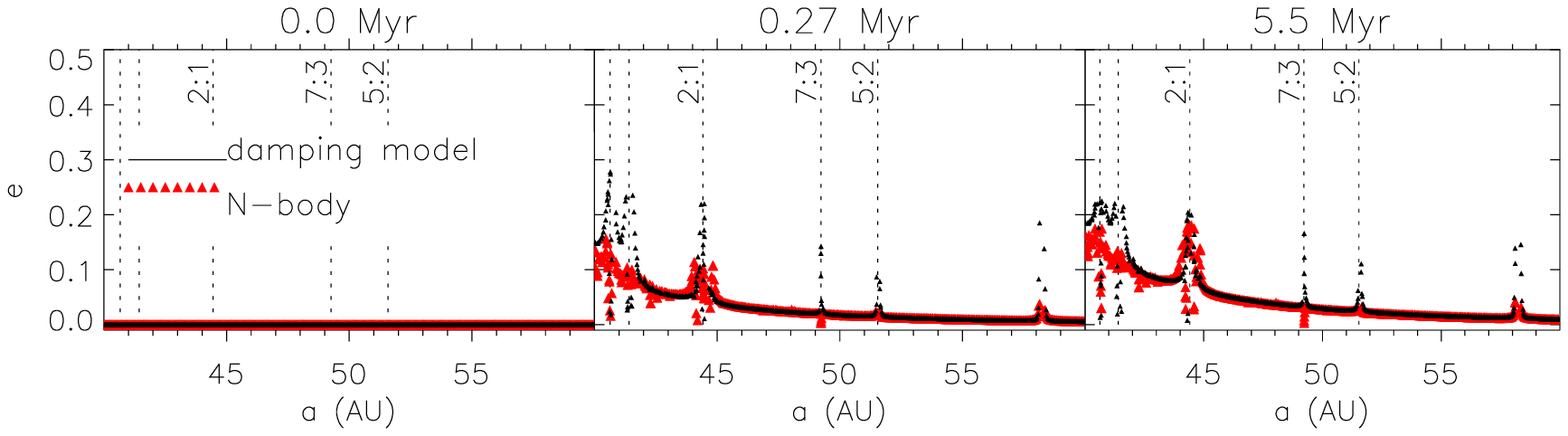}
\includegraphics[width=\textwidth]{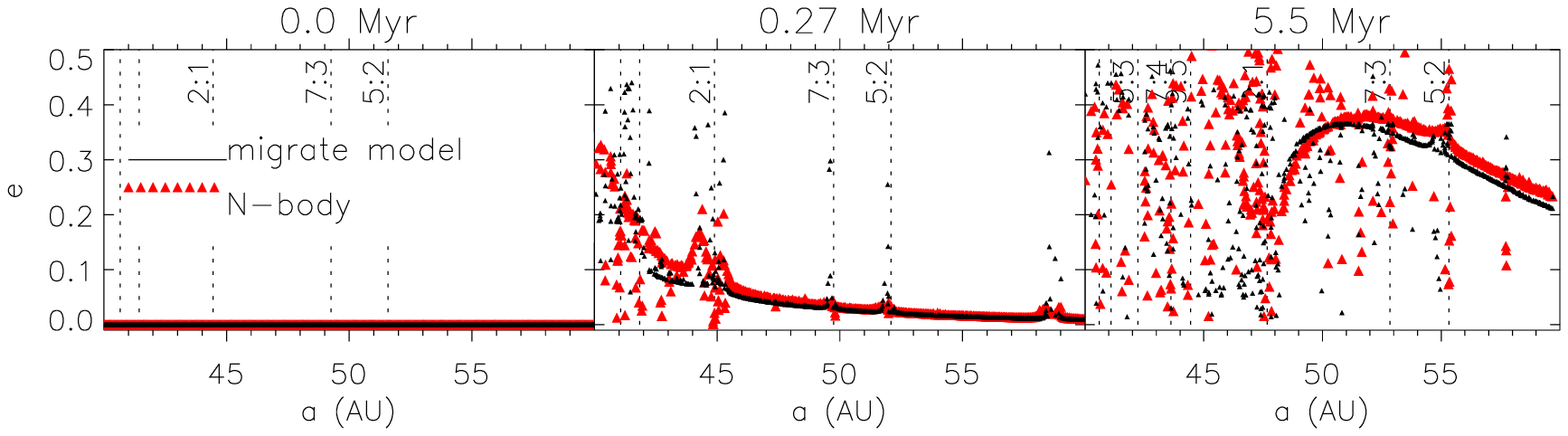}
\includegraphics[width=\textwidth]{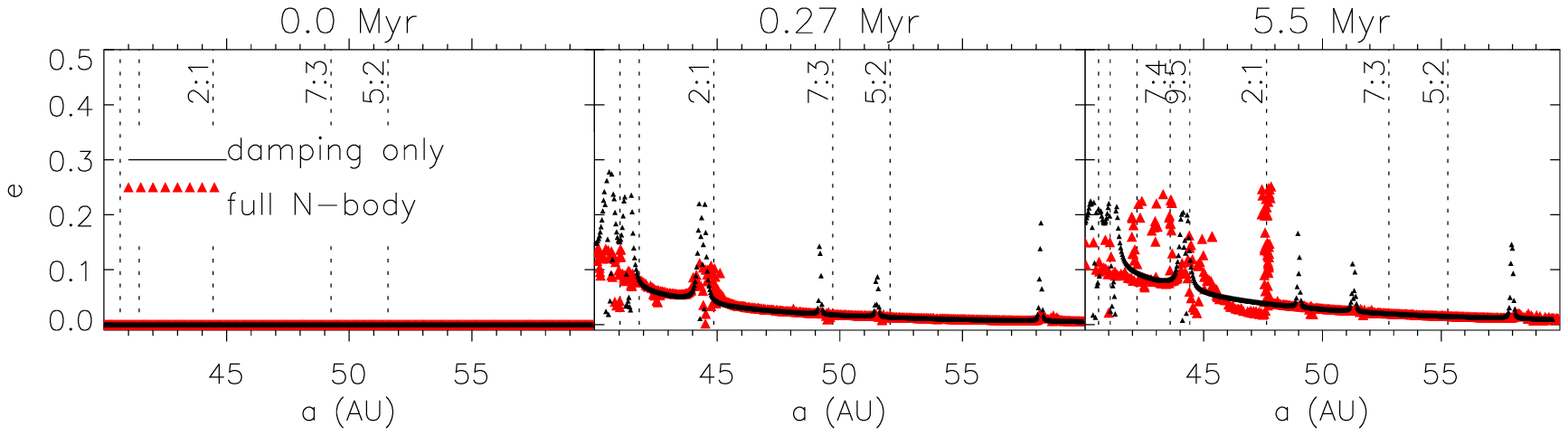}
\caption{Secular excitation of cold classicals when Neptune undergoes eccentricity damping and/or migration. In the top row, Neptune's eccentricity damps with $\tau_{\eN} =$ 0.3 Myr, and the planet does not migrate; secular theory, including damping (Equation \ref{eqn:effects}), is plotted in black. In the middle row, Neptune migrates outward on the timescale $\tau_{\aN} =$ 5 Myr, and its eccentricity does not damp. The secular theory, including migration (Equation \ref{eqn:effects}), is plotted in black. The scattered points in the final panel are objects that have undergone accelerated forcing near resonances as the resonances swept through. The bottom row displays a numerical integration including both eccentricity damping and migration, on the same timescales as above, but the analytical model overplotted in black is the same as in row 1.}\label{fig:demo}
\end{centering}
\end{figure*}

When Neptune migrates, $\alpha$ changes with time. In \citet{2012W}, we found that migration occurs in three regimes, relative to the eccentricity damping timescale $\tau_{\eN}$: fast, comparable, and slow. When Neptune's migration timescale $\tau_{\aN}$ is slow relative to the damping time, the secular evolution of the KBOs effectively takes place as if Neptune remains at its initial location. When Neptune's migration is fast relative to the damping time, the secular evolution of the KBOs effectively takes place as if Neptune were always at its final location. In the intermediate case, in which $\tau_{\eN} \sim \tau_{\aN}$, modeling the secular evolution at the location Neptune reaches after half a damping time is a fair approximation. In this work, we therefore use a fixed $\alpha$, which should be chosen according to these principles for a given evolution of Neptune. Figure \ref{fig:demo}, which we will describe after discussing our treatment of Neptune's eccentricity, provides an example showing that this approach is effective.

Before considering the impact of $\eN$ varying with time, we discuss the correction terms for Neptune's high eccentricity and for resonances. The correction terms for Neptune's high eccentricity do not change the form of the secular evolution. We have applied a correction term, $\frac{f_5}{f_2}\eN^2$, in our expression for $\gkbo$ (Equation \ref{eqn:extra}), and another such factor, $\frac{f_{10}}{f_{11}}\eN^2$, in $\gkbo'$ (Equation \ref{eqn:extrafactor}), which is another eigenfrequency. These terms are derived, the $f$ factors (which are of order unity) defined, and their necessity demonstrated in Appendix \ref{app:theory}. 
 
Proximity to resonance changes the secular frequency $\gkbo$ (Equation \ref{eqn:extra}), as described by \citet{1989M}. Orbital resonances greatly increase the secular forcing frequency because terms in the disturbing function that depend on the resonant angle can no longer be averaged over. The amplitude of the resonant correction term $\epsilon$ is defined in Appendix \ref{app:theory} and depends on how close the particle is to the location of mean-motion resonance. The frequency, $\delta \gkbo$, defined in Equation (\ref{eqn:extradelta}), depends on the order of the resonance $x$ and a constant $C$, of order unity, that is different for each resonance. See Figure \ref{fig:sec} for a demonstration of the resonance correction terms.

Finally, we turn to the impact of eccentricity damping, which alters $\eN$. Instantaneously, the KBO has the eccentricity components $h = e \sin \varpi =  e_{\rm free} \sin \phi + \bar{e}_{\rm forced} \sin \varpiN$ and $k = e \cos \varpi=  e_{\rm free} \cos \phi + \bar{e}_{\rm forced} \cos \varpiN$, where $\phi~=~\gkbo~t~+~\beta$. However, now the forced eccentricity vector ($\bar{e}_{\rm forced} \cos \varpiN, \bar{e}_{\rm forced} \sin \varpiN$) is changing. We have already accounted for the apsidal precession, but now the magnitude of the forced eccentricity vector is changing as well. When $e_{\rm forced}$ does not change, $e_{\rm free}$ is a constant determined by initial conditions. This remains true when $e_{\rm forced}$ evolves slowly compared to the secular forcing time of the KBO. Otherwise, $e_{\rm free}$ changes. Instead of allowing both $e_{\rm free}$ and $e_{\rm forced}$ to change with time, we use $e_{\rm free, 0}$ and define an ``effective" forced eccentricity, $| \bar{e}_{\rm forced}|$ (Equation \ref{eqn:extraforced}). The ``effective" forced eccentricity changes with time proportionally to $\eN$. Throughout the rest of the paper, we will refer to $e_{\rm free, 0}$ as $e_{\rm free}$ and the ``effective" forced eccentricity as $e_{\rm forced}$. The ``effective" forced eccentricity includes a factor $\sin ({\rm min} (g_{\gkbo} \tau_{\eN},\pi/2))$. When Neptune's eccentricity damping timescale is long compared to the KBO's secular oscillation period (i.e. $\gkbo \tau_{\eN} > \pi/2$), the particle's total eccentricity damps to its initial free eccentricity, $e_{\rm free, 0}$, and this factor is unity. However, when Neptune's eccentricity damps quickly, the particle's total eccentricity damps to a value near the eccentricity it reached after one damping time. The empirical correction factor allows us to model the particle's behavior without altering the form of the secular evolution. This empirical factor provides a match to the integrations (see Figure \ref{fig:demo}).

The modified secular theory, Equation (\ref{eqn:effects}), matches the integrations even when damping and migration are included (Figure \ref{fig:demo}).  Figure \ref{fig:demo} shows an example of a case in which $\tau_{\eN} = 0.3$ Myr is shorter than its migration timescale, $\tau_{\aN} =$ 5 Myr. (See Section \ref{subsec:short} for a description of how we implemented the damping and migration of Neptune's orbit.) In the top row, Neptune undergoes eccentricity damping from $e = 0.3$ at constant semimajor axis 28 AU. In the middle row, Neptune migrates from 28 AU to 30 AU on a timescale $\tau_{\aN} =$ 5 Myr, without its eccentricity damping. Unlike any other model in this paper, we use a time-dependent $\alpha(t)$ for the analytical model. In the bottom row, Neptune undergoes both damping and migration, but we plot again the same analytical model as in row 1, in which Neptune undergoes only eccentricity damping (no migration). The model over plotted in row 3, even though it in no way includes the effects of migration, matches well. When the migration is slow compared to the damping, the change in the secular frequency $\gkbo$ is negligible over the timescale during which Neptune's eccentricity is high. It is as if Neptune's eccentricity damps while Neptune remains at its initial $\aN$. Thus when the migration is slow compared to the damping timescale, we can model the KBOs' secular evolution as if Neptune's eccentricity damps while Neptune remains in place.

\subsubsection{Effects of other planets}
\label{subsec:otherplanets}
Our approach of modeling only Neptune, undergoing a range of orbital histories that would be caused by interactions with the other giant planets and with the solar system's planetesimal disk, is sufficient because the other planets primarily affect the KBOs only indirectly, through influencing Neptune. The influence on the Kuiper Belt of a single, apsidally precessing Neptune matches the influence of multiple planets in both integrations and theory.

\begin{figure}[htbp]
\begin{centering}
\includegraphics[width=.5\textwidth]{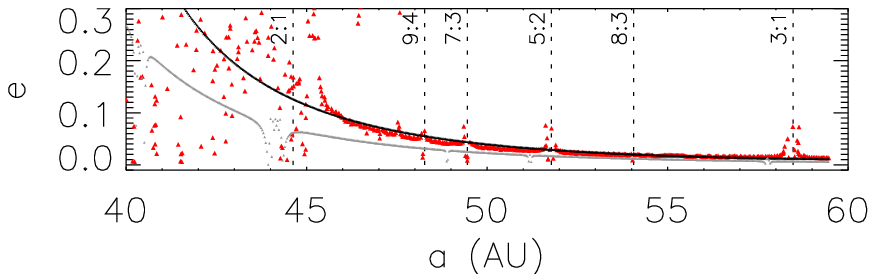}
\includegraphics[width=.5\textwidth]{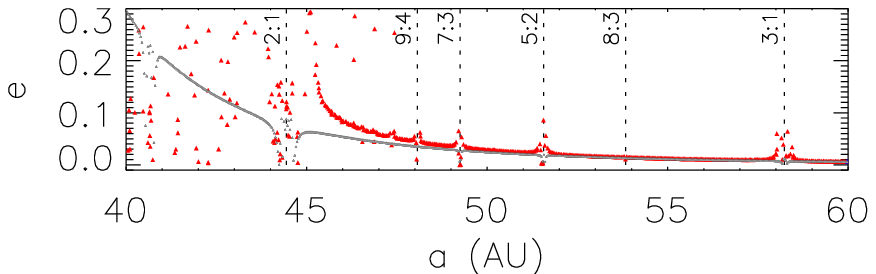}
\includegraphics[width=.5\textwidth]{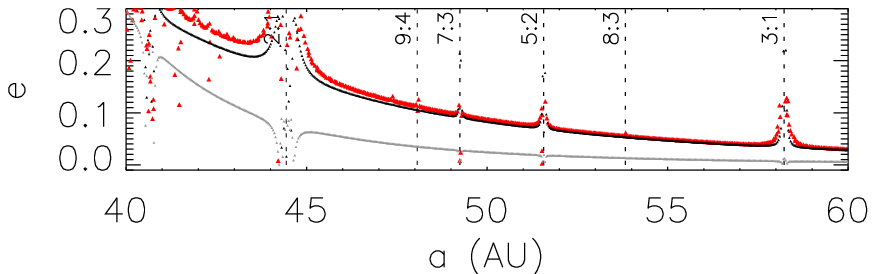}
\caption{Other giant planets affect the cold classicals indirectly by causing apsidal precession of Neptune. Top panel: snapshot (1.4 Myr) from an integration including four planets (with initial conditions $a$ = 5.2 AU, $e = 0.05$, $i=1.3^\circ$ for Jupiter, $a$ = 9.54 AU, $e = 0.06$, $i=2.49^\circ$ for Saturn, and $a$ = 16 AU, $e = 0.05$, $i=0.773^\circ$ for Uranus). Middle panel: same for integration including just Neptune undergoing apsidal precession with a period of 1.6 Myr. Bottom panel: same for integration including Neptune not undergoing precession. The black line in the top plot is the first-order multi-planet secular theory (i.e. without the extra resonant correction terms or higher order $e_{\rm N}$ terms we included for the just-Neptune theory) \citep{2000M}. The gray line on each plot is the analytical expression from the middle panel for comparison, computed using single planet secular theory, including precession (Equation \ref{eqn:effects}). The black line in the bottom panel does not include precession. Note the scattering in the top two panels interior to the 2:1 resonance.}\label{fig:compare4}
\end{centering}
\end{figure}

An illustrative case is shown in Figure \ref{fig:compare4}. The initial conditions for the particles are the same as in Figure 8, and Neptune has $\aN = 28$ and $\eN = 0.2$. The top panel (four planets) and middle panel (just Neptune undergoing precession) are very similar, keeping the objects at lower eccentricities than in the bottom panel (just Neptune, no precession). Thus precession must be included in the parameter space exploration, and including precession successfully accounts for the influence of the other giant planets. We note that in the time of the snapshot (1.4 Myr), the particles have secularly evolved to have eccentricities large enough so that their orbit, at the proper orientation, could intersect Neptune's. In cases in which Neptune precesses (top two panels), the objects are scattered by Neptune as Neptune's orbit precesses to intersect the orbits of the KBOs with eccentricities above the scattering line. The cut-off is at 45 AU because, interior this location, particles are secularly evolving quickly due to their proximity to resonance and have thus reached high eccentricities, allowing them to scatter. In the final panel, the particles are not scattered because Neptune's orbit does not precess to intersect the orbits of the particles.

The other planets matter in that they affect Neptune, but their direct effect on the KBOs is negligible. Their main effect is to cause precession of Neptune. Interactions between Uranus and Neptune, if they closely approach the 2:1 resonance while Neptune's eccentricity is high, cause additional effects that we discuss in Section \ref{subsec:chaos}. These effects can also be modeled using only Neptune, with appropriate orbital variations.

Treating only Neptune reduces the number of free parameters, allowing a more thorough exploration of the restricted space. The constraints we will develop can be applied to more extensive models that include the other giant planets. See Appendix \ref{app:planets} for a mathematical discussion of how the full four-planet secular theory reduces to the just-Neptune case.

\subsubsection{Constraints from secular excitation of cold objects}
It follows from the expressions in Section \ref{subsec:refined} that the excitation of the cold classicals happens on timescales of millions of years (or shorter near resonances) with an amplitude and timescale that depend on Neptune's semimajor axis, eccentricity, eccentricity damping timescale, migration rate, and precession rate. Complementarily, hot objects scattered into the classical region can evolve to lower eccentricities on similar timescales.

The observations require that the cold classicals not be excited above $e > 0.1$ in the region $42.5$ AU $< a < 45$ AU, as demonstrated in Section \ref{sec:obs}. From Equations (\ref{eqn:effects}) and (\ref{eqn:extra}) derived in Section \ref{sec:theory}, it follows that the cold classicals will not be excited above $e > 0.1$ at a given location if (CONSTRAINT 1):

\begin{equation}
\label{eqn:c1}
\sin ({\rm min}(\gkbo \tau_{\eN},\pi/2)) |  \frac{\gkbo'}{\dotvarpiN- \gkbo } | \eN < 0.1
\end{equation}

Thus there are three possible regimes to ``preserve" an in situ population of cold classicals through Neptune's wild days:

\begin{enumerate}
\item The eccentricity of Neptune is small enough that the region's forced eccentricity, proportional to $\eN$, is below 0.1 (i.e.  $|\frac{\gkbo'}{\dotvarpiN- \gkbo } | \eN < 0.1$)
\item Neptune's periapse precesses quickly enough that the region's forced eccentricity, inversely proportional to $\dotvarpiN$, is below 0.1 (i.e.  $\dotvarpiN$ is large).
\item The eccentricity of Neptune damps quickly enough that the objects are not excited above 0.1 (i.e. $\tau_e$ is small).
\end{enumerate}

\subsection{Effects of post-scattering secular evolution on hot objects}

Hot objects that have been scattered into the classical region from the inner disk (Section \ref{subsec:scatter}) will undergo secular evolution when they arrive in the classical region. They will reach a given semimajor axis $a$ in the classical region with eccentricities between $1-r_{a,N}/a$ and $1-r_{p,N}/a$. Not all of these eccentricities are consistent with stable orbits over 4 Gyr (Figure \ref{fig:diagnosticall}). If a particle is scattered to a high eccentricity above the stable region, under certain conditions --- if $\tau_{\eN}$  is not too fast and Neptune imparts a forced eccentricity that is large enough relative to the particle's free eccentricity --- the particle can reach a region of long-term survival through secular evolution. 

\begin{figure}[htbp]
\begin{centering}
\includegraphics[width=.5\textwidth]{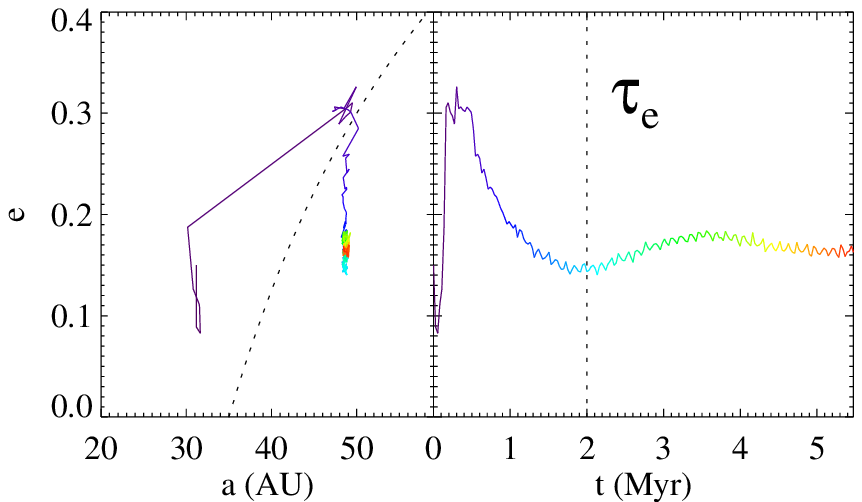} 
\includegraphics[width=.5\textwidth]{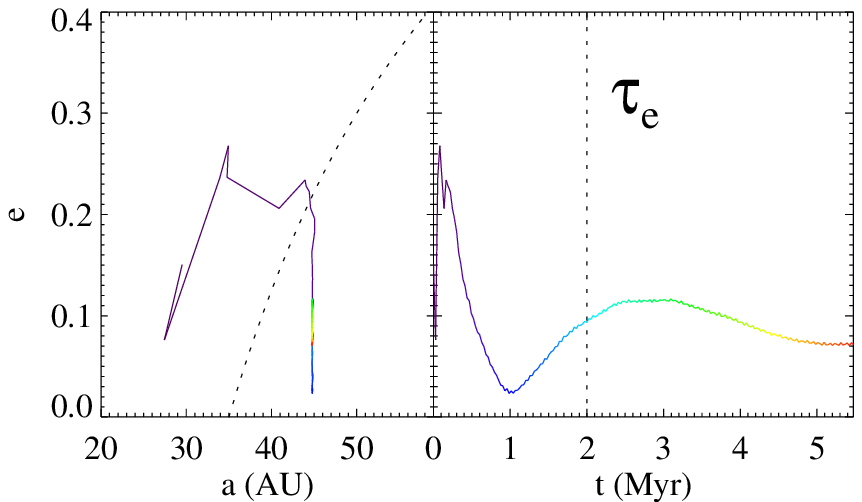}
\caption{Secular evolution can deliver hot objects into the classical region. Evolution of two example particles (top and bottom) for the first 6 Myr of an 4 Gyr integration resembling \citet{2008L} Run B. Left: path of objects in $(a,e)$ space. The color varies from purple (beginning of the integration) to red (6 Myr). The dashed line indicates the scattering line $q = 35$. Right: evolution of the particle's eccentricity versus time. The dashed line indicates one e-folding time, $\tau_{\eN} = 2$ Myr, for the damping of Neptune's eccentricity. \label{fig:examplescattered}}
\end{centering}
\end{figure}

In Figure \ref{fig:examplescattered}, we show two examples of KBOs that are scattered into the classical region from an integration resembling \citet{2008L} Run B. In this integration, Neptune begins with  $a_N = 28.9$ AU, $e_{\rm N} = 0.3$. Its eccentricity damps on a timescale of $\tau_e$ = 2 Myr, and it undergoes migration to 30.1 AU with $\tau_a = $ 10 Myr. Uranus begins with $a = 14.5$ and $e=0$ and also undergoes migration, to 19.3 AU, on the same timescale. Jupiter and Saturn begin with $a = 5.2 $ AU, $e =0.05$ and $a= 9.6$ AU, $e=0.05$ respectively and do not undergo migration or eccentricity damping. The integration includes 24,000 test particles, half of which (following \citealt{2008L}) begin in the region from 20-29 AU with $e = 0.2$ and half of which begin in the region from 29 - 34 AU with $e = 0.15$. The two example particles shown were among the group of particles found in the stable classical region in this integration after 4 Gyr and exhibit typical behavior. The particles are scattered into the classical region above the region of survival (Figure 3; below dotted line in right panel) but secularly evolve down into the stable region. After 4 Gyr, the particles remain in this location. Neither of these example objects is librating in an orbital resonance.

Therefore, Neptune's apoapse $r_{p,N}$ must be large enough so that the hot classicals reach the region of long-term survival either immediately or can evolve there before the eccentricity of Neptune damps, after which the eccentricity of the particle is frozen.

However, the post-scattering evolution depends strongly on $\varpi - \varpiN$. Particles are not actually scattered to orbits with independent, random $\varpi - \varpiN$.  This is because, by definition, after each scattering, the particle's new orbit fulfills the condition that at $\theta$, the angle at which the orbit of the particle and the orbit of Neptune intersect, $r = r_N$:
\begin{equation}
\label{eqn:scatter}
\frac{a(1-e^2)}{1+e\cos(\theta-\varpi)} = \frac{\aN(1-\eN^2)}{1+\eN\cos(\theta-\varpiN)}
\end{equation}

This is important because once the particle is scattered, it begins to undergo secular oscillations and the initial phase of the oscillation, $\beta_0$ (Equation \ref{eqn:effects}), depends on $\varpi - \varpiN$. 

Since in the classical region, most of the particle's orbit is outside of Neptune's orbit, the orbits will intersect close to the particle's periapse, i.e. the interior part of its orbit, so $\theta \approx \varpi$. When $\theta = \varpi$, Equation (\ref{eqn:scatter}) simplifies to:
\begin{equation}
\label{eqn:scatter0}
q = a (1-e) = \frac{\aN(1-\eN^2)}{1+\eN\cos(\varpi-\varpiN)}
\end{equation}
and thus $\varpi$ maps exactly to the particle's post-scattering periapse $q$. For example, particles scattered to the minimum $q = r_{p,N}$ have $\varpi- \varpiN = 0 $, while particles scattered to the maximum $q = r_{a,{\rm N}}$ have $\varpi - \varpiN = \pi $. Particles scattered to an intermediate $q = \aN(1-\eN^2)$ have $\varpi- \varpiN = \pm \pi/2 $.

\begin{figure}[htbp]
\begin{centering}
\includegraphics[width=.5\textwidth]{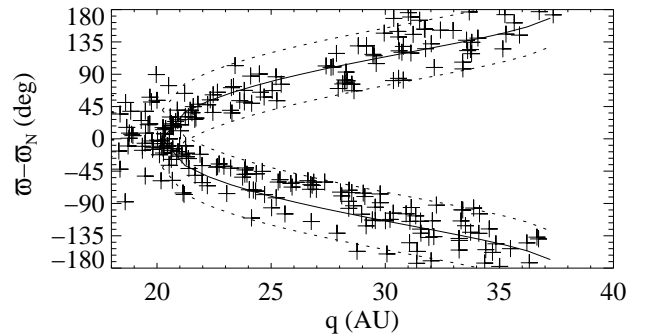}
\caption{Longitude of periapse relative to Neptune $\varpi - \varpiN$ of a particle's orbit after being scattered by Neptune into the region $a > 40$ AU maps to the particle's new periapse distance $q$. The particles from the integration are plotted as pluses and the solid line is Equation (\ref{eqn:scatter0}). The dashed lines are the solid line shifted by $\pm 40^\circ$.} \label{fig:scatterang}
\end{centering}
\end{figure}

In Figure \ref{fig:scatterang}, we plot the longitude of periapse relative that of to Neptune $\varpi - \varpiN$ versus periapse $q$ of test particles that were scattered in an integration we performed. The $\varpi - \varpiN$ is from the first (3000 yr) timestep after the particle's scattering. The integration lasted for 1 Myr and included Neptune, with $\aN = 30$ AU and $\eN = 0.3$, and 11,600 test particles evenly-spaced in semimajor axis from 29-34 AU, with $e = 0.15$. The $\varpi - \varpiN$ values are well-matched by Equation (\ref{eqn:scatter0}).

The secular evolution of the particle after its scattering will depend on its semimajor axis, eccentricity, and longitude of pericenter relative to Neptune's (Equation \ref{eqn:sec}). In particular, from Equation (\ref{eqn:sec}), it follows that the initial rate of change of the particle's eccentricity depends on $\varpi- \varpiN$. By calculating the particle's total eccentricity $e = \sqrt{h^2+k^2}$, differentiating with respect to time, and evaluating the time derivative at $t=0$, we find:
\begin{equation}
\label{eqn:edot}
\dot{e} (t = 0) = - \eN  \sin ( \varpi - \varpiN) \alpha b_{3/2}^{(2)}(\alpha) \frac{\mN}{m_\bigodot}   \frac{n}{4} 
\end{equation}
An analogous expression follows from Equation (\ref{eqn:effects}).

Along a scattering line $q$, the KBOs do not have random $\varpi$ but $\varpi$ close to the value dictated by Equation (\ref{eqn:scatter0}). Thus the particles scattered to the minimum $q = r_{p,N}$, which have $\varpi - \varpiN = 0$, and maximum $q = r_{a,{\rm N}}$, which have $\varpi- \varpiN = \pi $, are turning over in their secular evolution cycles $(\dot{e} = 0)$. However, particles scattered to an intermediate $q= \aN(1-\eN^2)$, which have $\varpi - \varpiN =\pi/2$, will be decreasing in eccentricity at the maximum rate in the cycle.

Moreover, the particle's free eccentricity also depends on $\varpi - \varpiN$:
\begin{equation}
\label{eqn:efree}
e_{\rm free}^2 = \bar{e}_{\rm forced}^2+(e(0))^2  - 2 e(0) \bar{e}_{\rm forced} \cos (\varpi-\varpiN)
\end{equation}
where $e(0)$ is the KBO's eccentricity at $t=0$. 

Thus $\varpi- \varpiN$ sets not only the particle's initial phase in its secular evolution cycle but also the amplitude of its free eccentricity. The particles with the phase to achieve the lowest possible total eccentricity ($e_{\rm free}$ close to $e_{\rm forced}$) may not initially be evolving downward in their cycle. Therefore, in order to calculate the minimum time for a particle to reach the stable region we consider all values of $\varpi- \varpiN$.

Thus particles with $\varpi- \varpiN = 0 $ and $\varpi - \varpiN = \pi$ have the smallest and largest free eccentricities respectively, while those with $\varpi - \varpiN =\pm \pi/2 $ have an intermediate value. Unfortunately for particles trying to reach low eccentricities, the ones with largest free eccentricity ($\varpi-\varpiN = \pi$) are initially going up (i.e. $e$ is increasing).

From numerical integrations (Figure \ref{fig:scatterang}), it appears that the maximum deviation in $\varpi-\varpiN$ as a function of $q$ from Equation (\ref{eqn:scatter0}) is $\pm 40^\circ$ for conditions relevant to Neptune and the classical region KBOs (due to the fact that $\theta$ in Equation (\ref{eqn:scatter}) is not always exactly $\varpi$). In setting the initial conditions for the secular evolution of scattered particles, we employ this mapping between $\varpi-\varpiN$ and the particle's $q$, including the uncertainty.

Thus the criterion for delivering the hot classicals developed in Section \ref{subsec:features} requires that:

CONSTRAINT 2: Neptune's apoapse is large enough ($r_{a,{\rm N}} > 34$ AU) so that particles are immediately scattered into the stable region, or Neptune imparts an $e_{\rm forced}$ large enough relative to $e_{\rm free}$ so that it is possible for particles with semimajor axes in the range 42.5-47.5 AU to evolve to $q > 34$ AU in less than $\tau_{e,N}$.

This constraint ensures that it is possible for at least some hot objects in the region from 42.5 to 47.5 AU to be delivered into the region of longterm stability.

\subsection{Accelerated secular forcing near resonances}
\label{subsec:restheory}

When two bodies are near resonance, the secular eccentricity forcing happens on a much faster timescale. This effect has been recognized as the cause of Saturn and Jupiter's fast precession, which \citet{1897H} attributed to their period ratios being near 5:2. More recently, \citet{2011M} recognized that the 2:1 resonance also contributes to Jupiter and Saturn's fast precession and \citet{1989M} identified the classical Uranian satellites' proximity to resonance as the cause of their deviation from their predicted ephemerides.

For the cold objects, the accelerated secular forcing near resonance quickly excites the eccentricities of these objects (as seen in Figure \ref{fig:sec}) , disrupting the confinement of the cold population. Near resonance, the correction term, $\epsilon \delta \gkbo $, to $\gkbo$ is large (Equation \ref{eqn:extra}). Thus the secular frequency $\gkbo$ is very high. If resonances overlay the cold classical region at early times, Neptune's eccentricity would have to damp on unrealistically short timescales to fulfill CONSTRAINT 1 (Equation \ref{eqn:c1}).  As shown in Figure \ref{fig:demo}, the objects near resonance remain dynamically disrupted even after Neptune's eccentricity damps. Since the cold objects in the region $42.5$ AU $< a < 45$ AU are confined to low eccentricities (Section \ref{subsec:features}), they cannot have been excited by accelerated secular forcing near resonance while Neptune's eccentricity was high.

CONSTRAINT 3: Resonances cannot overlie the region $42.5$ AU $< a < 45$ AU while Neptune's eccentricity is high. This constraint is a special case of CONSTRAINT 1 (Equation \ref{eqn:c1}) and is quantified in Section \ref{subsec:coldres}.

\begin{figure}[htbp]
\begin{centering}
\includegraphics[width=.5\textwidth]{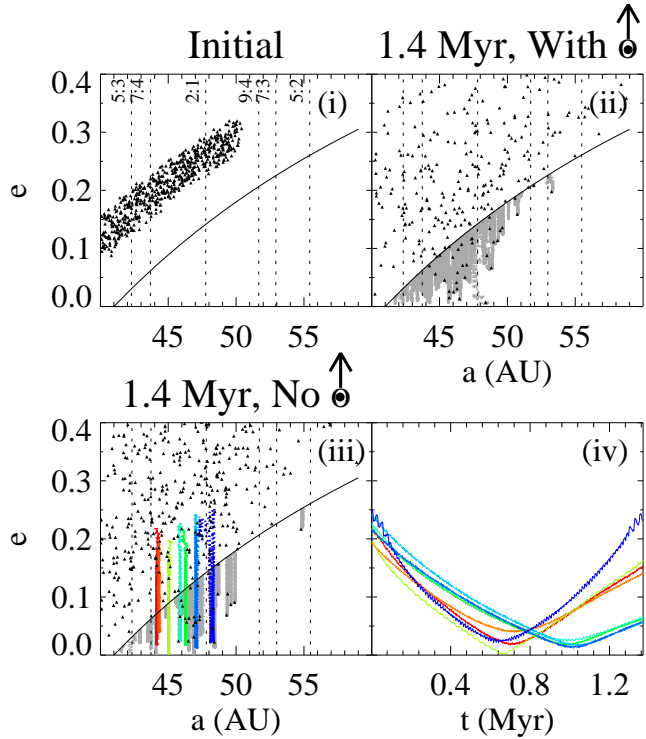}
\caption{Objects scattered into the classical region undergo re-scattering and secular evolution. Panel (i): Initial conditions of the particles. Panel (ii): Positions of the particles (black circles) after 1.4 Myr (compare to the middle row of \citet{2008L}, Figure 3) in an integration including all four giant planets. The gray is the cumulative region visited by the particles. Panel (iii): Same for an integration without Uranus. The colors show the paths of a few selected particles. Panel (iv): Eccentricity oscillations for particles corresponding to panel (iii).}\label{fig:resdemo2}
\end{centering}
\end{figure}

For the hot objects, accelerated secular forcing near resonance can drive down their eccentricities once they have been scattered into the classical region. Figure \ref{fig:resdemo2}, inspired by Figure 3 of \citet{2008L}, shows two example integrations of particles beginning at large eccentricities and evolving down to smaller eccentricities. The initial conditions (panel i) match those in \citet{2008L}, Figure 3 (their top left panel). The snapshots in panels (ii) and (iii) are after 1.4 Myr. The integration in panel (ii) includes all four giant planets. Neptune has initial conditions $\aN = 30$ AU and $\eN = 0.2$, and the other planets have their modern orbital elements. The integration in panel (iii) is the same except without Uranus. The particles are undergoing secular evolution, as demonstrated by their vertical paths in $(a,e)$ space ($a$ does not change under secular evolution) and their eccentricity oscillations (panel iv). If Uranus is present near the 2:1 resonance with an eccentric Neptune, the evolution of the particles is chaotic (panel ii), as we will discuss in more detail in Figure \ref{subsec:chaos}. For objects capable of reaching low eccentricities through secular evolution (Equation \ref{eqn:efree}), proximity to resonance significantly decreases the delivery time. Fast secular evolution near resonances can also assist in capturing objects into resonance, in addition to the previously identified mechanisms of chaotic capture \citep{2008L} and smooth migration \citep{1995M}.

\subsection{Chaotic sea: no additional constraints}
\label{subsec:chaos}

When Neptune's eccentricity is large, the resonances are widened and potentially overlap in what \citet{2008L} describe as a ``chaotic sea." \citet{2008L} argue that this region extends to the 2:1 resonance when Neptune's eccentricity is $e_{\rm N} > 0.15$. We have investigated the circumstances for chaos and reached several conclusions, which we will state and then justify:

\begin{enumerate}
\item Even when widened by Neptune's high eccentricity, the resonances between the 5:3 and 2:1 do not overlap except for particles at high eccentricities.
\item Variations in Neptune's semimajor axis on timescales of order a KBO libration time combine with widened resonances to cause a chaotic region.
\item The existence and extent of the chaotic region depend on the details of Neptune's interactions with Uranus.
\item The potential for chaos does not impose immediate additional constraints beyond those described in the previous subsection (Section \ref{subsec:refined}). We will show that we can still model just Neptune, taking into account its potential evolution under the influence of the other giant planets.
\end{enumerate}

\subsubsection{Just Neptune: no chaotic sea}

\begin{figure}
\begin{centering}
\includegraphics{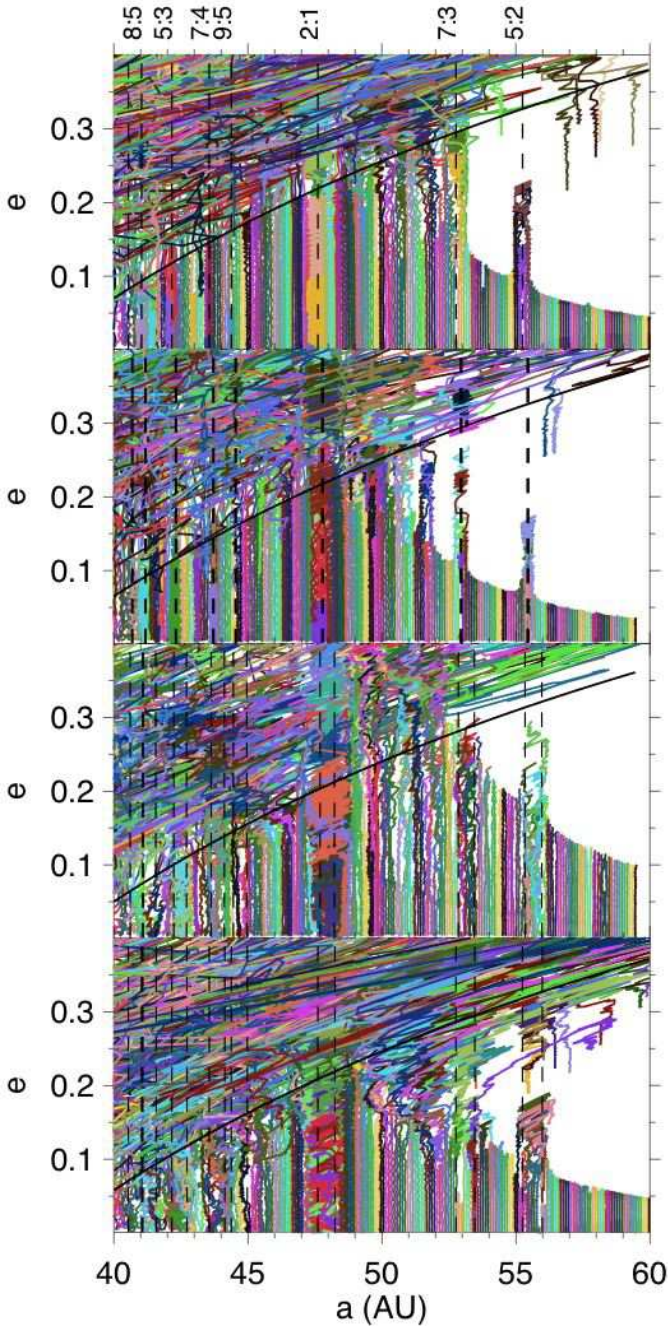}
\caption{Paths of particles in $(a,e)$ space provides a qualitative probe of the existence of a chaotic sea; the solid line indicates the scattering line. The locations of the resonance centers are plotted as dashed lines; in the case in which Neptune's semimajor axis changes (i.e. in rows 3 and 4), the minimum and maximum centers are plotted. We integrated 200 massless test particles, each starting with $e=0$, for 6 Myr under the influence of a subset of planets. Neptune, present in each integration, begins with $\aN = 30, \eN = 0.2$. Row 1: the top panel is an integration with just Neptune, precessing with a period of 4 Myr. The chaotic sea is not present: particles evolve secularly upward in $(a,e)$ space until they reach the scattering line. Row 2: the behavior of the particles in this integration, which also includes Jupiter and Saturn at their current locations, is qualitatively similar: no chaotic sea. Row 3: in this integration --- which includes Jupiter, Saturn, and Uranus at their current locations --- the chaotic sea appears, extending up to about 45 AU, just past the 9:5 resonance. Instead of a straight upward path, the particles move from left to right as well. Row 4: the chaotic sea is also present in this integration, which includes just Neptune but with its semimajor axis oscillating with a period of 12,400 yr and amplitude 0.2 AU, as it does under the influence of Uranus in row 3. Values for $a$ and $e$ are in barycentric coordinates.}\label{fig:chaoshappens}
\end{centering}
\end{figure}

In addition to examining each particle individually, a qualitative way to distinguish between: 1) a chaotic sea, and 2) particles secularly evolving until they are scattered, is to plot the individual paths of a collection of particles through $(a,e)$ space (Figure \ref{fig:chaoshappens}). First we consider an integration that includes only Neptune (row 1). Each starting at $e=0$, the particles move straight upward vertically in $(a,e)$ space until they reach the scattering line (solid black line, $e = 1 -r_{a,{\rm N}} /a$), rather than moving horizontally and vertically as they would in a chaotic sea.

In integrations including just Neptune, with or without apsidal precession, the eccentricity of an initially cold particle grows secularly until its orbit crosses Neptune's at $e > 1-  r_{a,{\rm N}}/a$. For a particle at $42.5$ AU, when $\aN = 30$ AU and $\eN = 0.2$, this threshold is $e >$ 0.15. After reaching this threshold, the particle then undergoes scattering events. Even particles near resonances evolve secularly, with the increased secular frequency defined by Equation (\ref{eqn:extra}). These particles appear separated from the coherent excitation of the other cold particles (Figure \ref{fig:sec}) because they undergo secular evolution so quickly and because the secular oscillation rate depends quite steeply on semimajor axis near resonance. It appears that, for particles that begin at low eccentricities, the resonances do not overlap even when $\eN = 0.2$. Therefore the constraints from \ref{subsec:refined} hold in the just-Neptune case, even when the planet is precessing.

\subsubsection{Neptune, Jupiter, and Saturn: no chaotic sea}
\label{subsubsec:noura}

Adding Jupiter and Saturn in their current configurations does not generate a chaotic sea (Figure \ref{fig:chaoshappens}, row 2). The particles continue to move upward in $(a,e)$ space until they reach the scattering line. The behavior is not qualitatively different from the just-Neptune case. 

\subsubsection{Neptune and Uranus: chaotic sea}

However, when Uranus is added on its current orbit, a chaotic sea appears in the classical region, extending up to the 9:5 resonance (Figure \ref{fig:chaoshappens}, row 3). In the chaotic regime, individual particles exhibit chaotic jumps in their eccentricity. Some cross from one resonance to another. They move horizontally, as well as vertically, in $(a,e)$ space. 

Why does adding Uranus create the chaotic sea? Neptune and Uranus exhibit anti-correlated variations in their semimajor axes associated with proximity to their 2:1 resonance. For the configuration considered here, the periodicity of this variation is about $10^4$ years, and the amplitude for Neptune is about 0.2 AU. This timescale is of order the typical libration time of a resonant KBO in the classical region. We performed additional integrations in which we modified \emph{Mercury6} to turn off the gravitational interaction between each KBO and any planet except Neptune. The behavior was qualitatively the same as that shown in Figure \ref{fig:chaoshappens}, row 1, suggesting yet again that the other giant planets only affect the KBOs indirectly through their impact on Neptune's orbit. We performed integrations that include just Neptune, no other giant planets, with its semimajor axis oscillating with a period of 12,400 yr and amplitude 0.2 AU. The behavior was the same as in the four planet case, and the chaotic sea was present (Figure \ref{fig:chaoshappens}, row 4). Evidently this strong periodicity in Neptune's orbital variations, which is driven by Uranus and which moves the locations of the resonances on timescales of order a KBO libration time, causes the chaotic sea.

We note that in the case in which Uranus is excluded (Section \ref{subsubsec:noura}, Figure \ref{fig:chaoshappens}, row 2), Neptune's semimajor axis is perturbed on orbital timescales (165 years) by Saturn and Jupiter, with an amplitude of 0.02 AU. This small-amplitude perturbation in $\aN$, on a timescale two orders of magnitude shorter than the resonant timescale of KBOs, does not create a chaotic sea. The chaotic sea appears to be limited to particular dynamical histories in which Neptune and Uranus are strongly interacting through the 2:1 resonance. A thorough exploration of these histories are beyond the scope of this paper, but may provide additional constraints.

The chaotic sea appears to extend to 45 AU, just past the 9:5 resonance. This region between the 3:2 and 9:5 resonances is already forbidden to overlie today's cold classical region because the secular precession rates are extremely fast there (Section \ref{subsec:restheory}), so --- in the case of the cold classicals --- the potential for chaos adds no additional constraints. We notice that 45 AU coincides with the current edge of the cold classical region. In our interpretation, this is a coincidence. We also note that \citet{2008L} found that the chaotic sea extended to the 2:1. Because the chaos depends on the interactions between Uranus and Neptune, we expect that this difference may be due to different initial conditions for Uranus. Moreover, their interpretation that the chaotic sea extends to the 2:1 is based on the fact that the cumulative region visited by the particles extends to the 2:1 \citep[Figure 3]{2008L}. An alternative interpretation is that chaotic sea extends only to the 9:5 but that particles beyond this location are close enough to either the 9:5 or the 2:1 resonance to quickly reach low eccentricities through secular evolution, which is faster in regions near resonances.

Our constraints based on secular evolution (Section \ref{subsec:refined}) are conservative. The chaotic sea cannot revive a region of parameter space which we have excluded, but it can rule out additional regions. If oscillations in Neptune's semimajor axis caused by interactions with Uranus are large enough, the chaotic sea may extend beyond the 9:5 resonance, which would impose additional constraints. For example, if the chaotic sea extended to the 2:1 resonance, the region of parameter space for Neptune with $\eN > 0.1$ between 28-29 AU (which we will demonstrate is viable in Section 5.1, Figure \ref{fig:ccold}), in which the cold classical region is sandwiched between the 9:5 and the 2:1 resonance, would no longer preserve the low eccentricities of cold classicals.

Another effect of the oscillations in Neptune's semimajor axis is to effectively widen the resonances. This effect could potentially cause more KBOs at the edge of, but not within, the chaotic sea (for example, KBOs just beyond the 2:1 resonance) to experience fast secular evolution due to proximity to orbital resonance (Section \ref{subsec:restheory}). Thus even more parameter space could be ruled out. We do not explicitly take this into account because the oscillations in Neptune's semimajor axis depend on the particular configuration of Neptune and Uranus.

Our constraints are a starting point for more extensive integrations, which will require careful consideration of the interactions between Neptune and Uranus. We note that, for illustrative purposes, we have used the current semimajor axis and eccentricity of Uranus and current semimajor axis of Neptune in these explorations, but the existence and extent of the chaotic sea depends on their particular orbital configuration --- especially their proximity to the 2:1 resonance --- during Neptune's wild days.

\subsection{Summary}
\label{subsec:sumtheory}
By modeling the four dynamical processes that result from Neptune's high eccentricity ---  scattering (Section \ref{subsec:scatter}), secular forcing (Section 4.2-4.3), accelerated secular forcing near resonances (Section \ref{subsec:restheory}), and a chaotic sea (Section \ref{subsec:chaos}) --- we have translated the conservative criteria imposed by the observed eccentricity distributions of the hot and cold classicals (Section \ref{subsec:features}) into the following constraints:

\begin{itemize}
\item Neptune's apoapse must be large enough to deliver hot objects to the longterm-stable classical region immediately ($r_{a,{\rm N}} > 34$ AU) or $e_{\rm forced}$ must be large enough (relative to $e_{\rm free}$) to evolve the particle's $e$ to $< 0.3$ in less than Neptune's eccentricity damping time $\tau_{\eN}$
\item The final value for the eccentricities of planetesimals in the region from $42.5$~AU~$<~a~<~45$~AU must be less than $e = 0.1$: $ \sin ({\rm min} (\gkbo \tau_{\eN},\pi/2))  \frac{\gkbo'}{\dotvarpiN- \gkbo } \eN < 0.1$ (or the forced eccentricities must be kept below 0.1 by fast precession).
\item Resonances cannot overlie the region $42.5$ AU $< a < 45$ AU while Neptune's eccentricity is high.
\end{itemize}

Thus Neptune's eccentricity must be high in order to deliver hot objects to the classical region and yet will disrupt the cold objects quickly unless Neptune's high eccentricity damps quickly or the planet's orbit apsidally precesses quickly. In all cases, mean motion resonances with Neptune cannot overlie the region $42.5$ AU $< a < 45$ AU while Neptune's eccentricity is large.

\section{Results: constraints on Neptune's dynamical history}
\label{sec:constraints}

Applying the constraints developed in Section \ref{sec:process}, we place constraints on parameters of Neptune during its dynamical history. First, we consider separately which parameters of Neptune preserve an in situ cold population (Section \ref{sec:cold}) and which allow delivery the hot classicals (Section \ref{sec:hot}). In Section \ref{sec:combo}, we combine those constraints and determine which parameters of Neptune allow the planet to simultaneously preserve the cold classicals while delivering the hot classicals. In Section \ref{subsec:combointerpret}, we interpret these parameter constraints in light of Neptune's full dynamical history. We present example integrations illustrating the constraints in Section \ref{subsec:example}.                                                                

The combined constraints will offer answers to the following questions about Neptune's dynamical history:
\begin{itemize}
\item Could Neptune have been scattered to a high-eccentricity orbit?
\item If so, how quickly did dynamical friction damp Neptune's eccentricity?
\item How far did Neptune migrate in the protoplanetary disk?
\item If both damping and migration occurred, what were their relative timescales?
\end{itemize}

\subsection{Regions of parameter space that keep cold objects at low eccentricities}
\label{sec:cold}

In Section \ref{subsec:nores}, we identify which regions of parameter space fulfill CONSTRAINT 1 (Equation \ref{eqn:c1}), preserving the cold classicals below $e < 0.1$ (as summarized in Section \ref{subsec:sumtheory}), without including the effects of orbital resonances or precession. In Section \ref{subsec:coldres}, we incorporate CONSTRAINT 3, the effects of orbital resonances. Finally, in Section \ref{subsec:precess}, we consider the special case of fast precession.

\subsubsection{Constraints on Neptune's eccentricity and damping time}
\label{subsec:nores}

\begin{figure}[htbp]
\begin{centering}
\includegraphics[width=.5\textwidth]{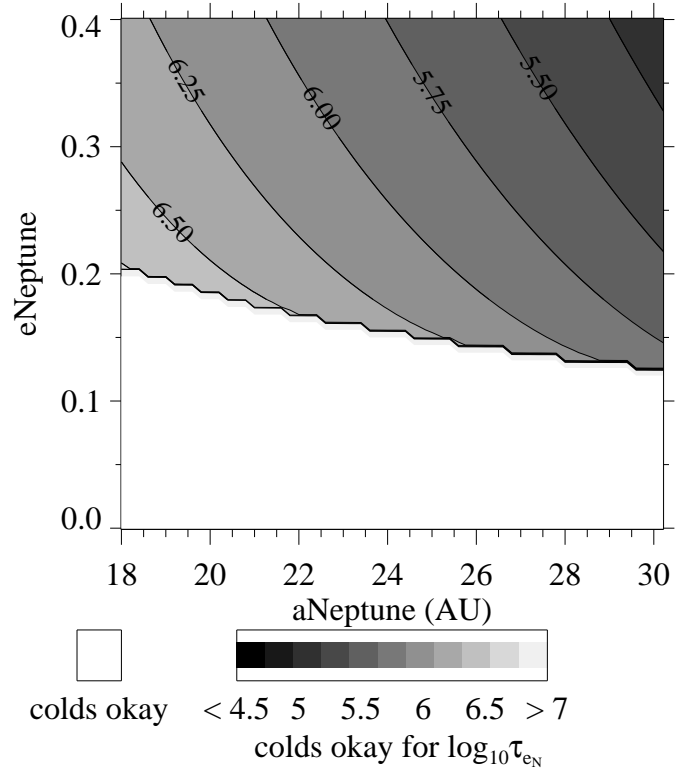}
\caption{Constraints on Neptune's parameters to preserve the forced eccentricities of KBOs below $e_{\rm forced} = 0.1$ in the region from 42.5 to 45 AU, where $a_{\rm Neptune} = \aN$ and $e_{\rm Neptune} =\eN$ are spelled out for clarity. The white region indicates parameters of Neptune that keep $e_{\rm forced} < 0.1$, no matter how long the damping time. The gray regions are contours of the maximum allowed $\log_{10} \tau_{\eN}$ (where $\tau_{\eN}$ is in years), neglecting orbital resonances and assuming Neptune's orbit has zero precession, to fulfill CONSTRAINT 1 (Equation \ref{eqn:c1}).} \label{fig:ccoldnotres}
\end{centering}
\end{figure}

We begin by identifying regions of parameter space where, for a given semimajor axis of Neptune $\aN$, Neptune's eccentricity $\eN$ is small enough or its damping time $\tau_{\eN}$ is short enough to avoid excessively exciting the cold classicals. In this subsection, we neglect the effects of resonances and assume zero precession of Neptune's orbit.  in Figure \ref{fig:ccoldnotres} reveals the two two main regions of parameter space. 

\begin{enumerate}
\item In the contoured region (high $\eN$), we plot contours of maximum eccentricity damping time as a function of $(\aN,\eN)$ that fulfill the criteria set by CONSTRAINT 1 (Equation \ref{eqn:c1}). We calculate the maximum eccentricity damping time by considering the equation in CONSTRAINT 1 for a particle at 42.5 AU, where the secular evolution is fastest (excluding resonances). This map of ($\aN,\eN$) illustrates constraints on how quickly Neptune's eccentricity must damp --- in order to avoid exciting the cold classicals above the level we observe --- when Neptune occupies a particular region of $(\aN, \eN)$ space. 
\item In the white region, Neptune's eccentricity is small enough that the forced eccentricity of the cold classicals never exceeds $e_{\rm forced} = 0.1$ and thus the damping time can be arbitrarily long. As shown in \citealt{2012W}, in the regime of slow damping, the particle's eccentricity damps to its free eccentricity which, for a KBO beginning with $e = 0$, is equal to $e_{\rm forced}$.
\end{enumerate}

We will present several more such plots throughout the paper. Note that we plot only eccentricities up to $e_{\rm N} = 0.4$, corresponding to a $r_{p,N} = 18$ AU for $\aN = 30 $ AU. Above this value, corrections for Neptune's high eccentricity beyond what we already included would be necessary, and one also worries about the orbit of Neptune crossing the orbit of Uranus, which is currently situated at 19 AU. However, the constraints we have developed could be considered for larger $\eN$.

The criteria in this subsection hold when the secular excitation times are not affected by proximity to resonance.

\subsubsection{Constraints on Neptune's dynamical history, including the effects of resonances}
\label{subsec:coldres}

\begin{figure}[htbp]
\begin{centering}
\includegraphics[width=.5\textwidth]{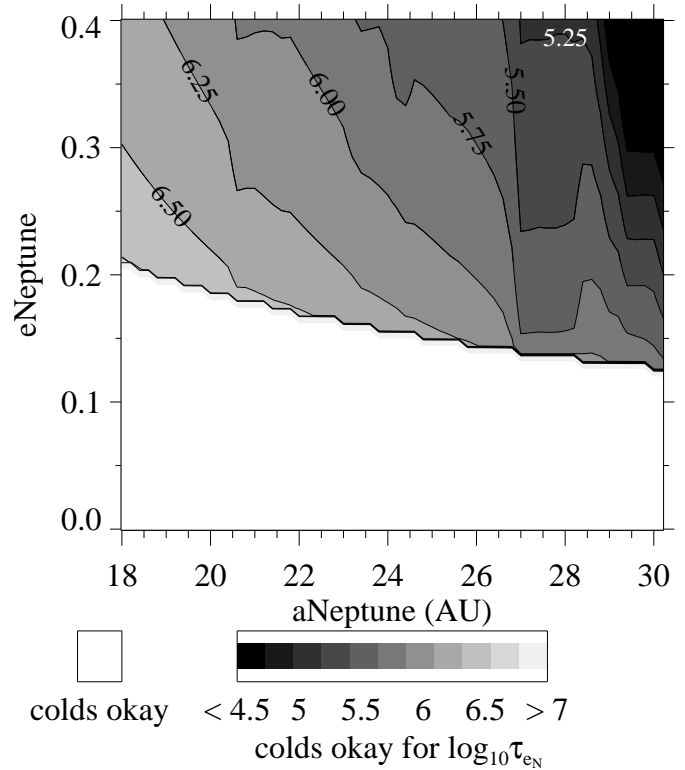}
\caption{Contours of $\log_{10} \tau_{\eN}$ (where $\tau_{\eN}$ is in years), from the constraint that the small bodies be preserved below $e < 0.1$ in the region from 42.5 to 45 AU, where $a_{\rm Neptune} = \aN$ and $e_{\rm Neptune} =\eN$ are spelled out for clarity. The white region indicates where Neptune's eccentricity is small enough that the forced eccentricity of the cold classicals never exceeds $e = 0.1$. Note that the dark regions are forbidden to Neptune; in these regions, the resonances overlie the cold classical region, exciting the cold classicals to high eccentricities on timescales less than $10^{4.5}$ years. The constraints in this plot correspond to the situation in which at least $80\%$ of the particles in the region from 42.5 to 45 AU are confined below $e < 0.1$.}  \label{fig:ccold}
\end{centering}
\end{figure}

In regions near orbital resonance with Neptune, KBOs undergo significantly faster secular evolution, as demonstrated in Section \ref{sec:theory}. Here we incorporate the resonance correction terms for the secular excitation times. in Figure \ref{fig:ccold}, analogous to Figure \ref{fig:ccoldnotres}, we plot contours of eccentricity damping time as a function of $(\aN,\eN)$ that fulfill the criteria set by CONSTRAINT 1 (Equation \ref{eqn:c1}) . We calculate the damping time of Neptune's eccentricity, $\tau_e$ for which 80$\%$ of initially cold objects in the region 42.5 AU $< a < 45$ AU remain below $e = 0.1$. Note the key difference between the two figures: in Figure \ref{fig:ccoldnotres}, $\tau_e$ varied smoothly through $(\aN,\eN)$ space, but in Figure \ref{fig:ccold}, there are dark regions where the eccentricity damping time is substantially reduced due to resonances overlying the classical region. 

The damping times are unrealistically short for parameters of Neptune for which resonances overlie the cold classical region. Neptune is unlikely to have spent substantial time with high eccentricity in these $\aN$ ranges. \citet{2007F} find that a Neptune-mass planet at 20 AU with an eccentricity of 0.3 undergoes eccentricity damping on $\tau_{\eN} =$ 0.6 Myr to 1.6 Myr (depending on whether Neptune's orbit intersects the planetesimal disk at pericenter/apocenter or at quadrature) if the surface density $\sigma$ of planetesimals is 1 g/cc. This is roughly the surface density needed to grow planetesimals in the region of Neptune \citep{1998K}. Since $\tau_{\eN} \propto \sigma^{-1},$ it is unlikely to be less than 0.1 Myr ($10^5$ yr), an order of magnitude faster than the $\tau_{\eN}$ calculated by \citet{2007F}. With resonances incorporated, certain regions cannot satisfy CONSTRAINT 1 (Equation \ref{eqn:c1}) in the zero precession case without un-physically low values of $\tau_{\eN}$. CONSTRAINT 3 is a qualitative statement of this result.  Thus by applying CONSTRAINT 1, including resonances, we recover CONSTRAINT 3.

\begin{figure}[htbp]
\begin{centering}
\includegraphics[width=.5\textwidth]{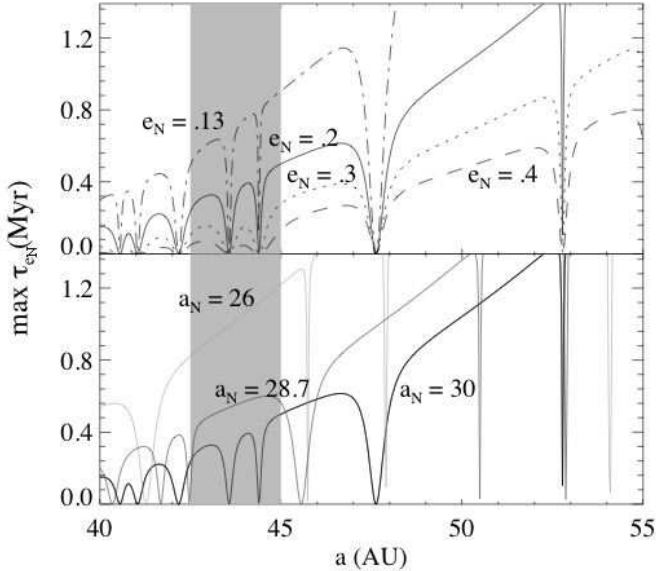}
\caption{Maximum eccentricity damping time $\tau_{\eN}$ (abbreviated $\tau_e$) to preserve the cold classicals at $e < 0.1$, as a function of a KBO's semimajor axis for illustrative parameters of Neptune. The gray region is the cold classical region ($42.5$ AU $< a < 45$ AU). Top panel: maximum eccentricity damping times for: $\aN = 30$ and $\eN =$ 0.4, 0.3, 0.2, and 0.13. Bottom panel: times for $\aN = 30, 28.7, 26$ for $\eN = 0.2$. }\label{fig:excite}
\end{centering}
\end{figure}

Incorporating the resonance correction terms, we plot the maximum $\tau_{\eN}$ as a function of the KBO's semimajor axis for several combinations of $\aN$ and $\eN$ (Fig \ref{fig:excite}). The maximum eccentricity damping time is larger for lower eccentricities, as shown for several illustrative values of $\eN$ in the top panel. The dips where $\tau_{\eN}$ approaches 0 correspond to regions of the Kuiper Belt near mean motion resonances where the secular excitation time is very short. The dips are wider when Neptune's eccentricity is higher. When these dips overlie the cold classical region from $42.5$ AU $< a < 45$ AU, the damping time requirement is un-physically short. In the bottom panel, we see how the resonances shift with Neptune's semimajor axis. At $\aN = 30$ AU, several resonances overlie the cold classical region. When $\aN = 28$ AU, the resonances are shifted interior and the cold classical region is sandwiched between two resonances, the 9:5 and the 2:1. For $\aN = 27.5 $ AU (not shown), the 2:1 resonance would be on top of the cold classicals. When $\aN = 26$ AU, all the major resonances are interior to the cold classical region. 

\subsubsection{Special regimes of fast precession}
\label{subsec:precess}

\begin{figure}[htbp]
\begin{centering}
\includegraphics[width=.5\textwidth]{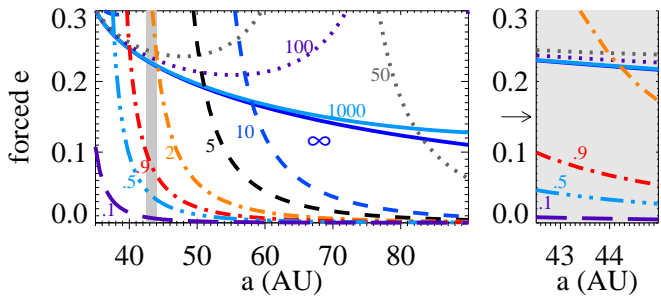}
\caption{Forced eccentricity vs semimajor axis for a range of precession rates, for $\eN = 0.3, \aN=30$. The forced eccentricity stays below 0.1 in the region  $42.5$ AU $< a < 45$ AU for $2\pi/\dot{\varpi} < 0.9$ Myr }\label{fig:precessexcite}
\end{centering}
\end{figure}

In this subsection, we explore the special case of fast precession. \citet{2011B} first suggested that if Neptune were to precess sufficiently quickly, the cold classical population would remain unexcited because fast precession lowers the forced eccentricity (Equation \ref{eqn:extra}). In Figure \ref{fig:precessexcite}, we plot the forced eccentricity as a function of the particle's semimajor axis for a range of precession rates when $\aN = 30$ AU, $\eN = 0.3$. Thus $\tau_{\eN}$ in the allowed region can be arbitrarily long if the precession period is sufficiently short. A precession period of 0.9 Myr keeps $e_{\rm forced} < 0.1$ in the cold classical region for these parameters of Neptune (compare to the current $g_8$ precession period of 2 Myr). However, an even faster apsidal precession period for Neptune may be required if not only secular evolution but scattering and/or chaos excites the cold classicals.

Unlike in the case of fast damping (Section \ref{subsec:coldscatter}), in the case of fast precession, scattering cold KBOs by Neptune may impose an important constraint. In the case of fast precession (and slow damping), even though the KBO's eventually evolve to $e_{\rm free}$ once Neptune's eccentricity damps, the KBO can reach a maximum value of $e_{\rm max} = 2e_{\rm forced}$ while $\eN$ is still high. Thus the KBO is vulnerable to being scattered. For example, for$\aN = 30$ AU, $\eN = 0.3$, a KBO at 42.5 AU can be scattered if it reaches $e > 0.08$. Thusthe KBO needs $e_{\rm forced} < \frac{1}{2} 0.08 = 0.04$ to avoid being scattered, requiring a fast precession period of 0.4 Myr for Neptune.

Chaos (Figure \ref{fig:chaoshappens}) can still excite the cold classicals even if their forced eccentricity is low. In this case, resonances cannot overlie the cold classical region if Neptune's semimajor axis oscillates strongly due to interactions with Uranus (Section \ref{subsec:chaos}). Therefore, if the chaotic sea is present, the dark regions in Figure \ref{fig:ccold} are still forbidden. 

\begin{figure}[htbp]
\begin{centering}
\includegraphics[width=.5\textwidth]{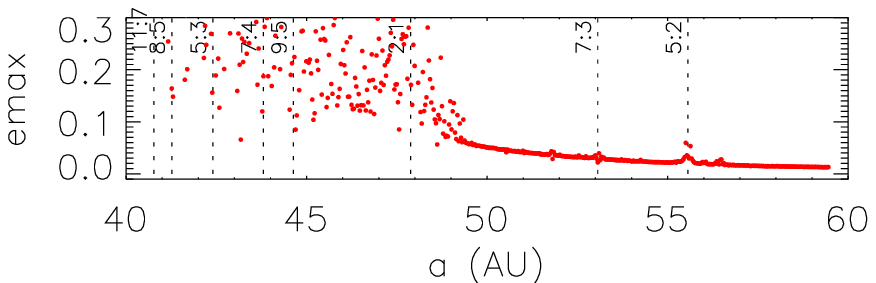}
\includegraphics[width=.5\textwidth]{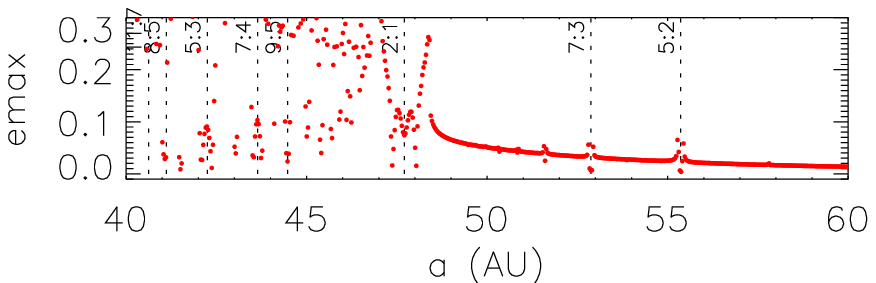}
\caption{If fast precession is caused by Neptune's interactions with Uranus, chaos can disrupt the confinement of the cold classicals. Top panel: maximum eccentricity reached by each particle during a 1.6 Myr integration including all four giant planets. Despite fast precession, objects interior to the 2:1 resonance are excited to high eccentricities. Bottom panel: maximum eccentricity when just Neptune precesses at the same rate (period of 1.2 Myr) as in the four-planet case.\label{fig:comparechaos}}
\end{centering}
\end{figure}

Figure \ref{fig:comparechaos} demonstrates how fast precession caused by Neptune's interaction with the other planets is unable to keep the cold objects at low eccentricities in resonance regions. We performed two integrations; in each of which Neptune has $\aN = 30.06$ and $\eN = 0.2$, and 600 test particles begin with $e = 0$. The first integration (top panel) includes the other three giant planets, on their current orbits; they cause Neptune to undergo apsidal precession with a period of 1.2 Myr. In the second integration, Neptune is forced to precess at this rate without any other planets included (see Section \ref{subsec:short} for details on the implementation). We plot the maximum eccentricity reached by each particle. In the top panel, chaos has dynamically disrupted particles that were preserved at low eccentricities by fast precession in the bottom panel (i.e. in the region interior to 45 AU).

Unfortunately, the most obvious configuration that causes fast precession --- Neptune and Uranus near their 2:1 resonance --- also causes large oscillations in Neptune's semimajor axis. The larger the oscillations in Neptune's semimajor axis, the wider the chaotic sea. Although \citet{2011B} suggest that a massive planetesimal disk could contribute to fast precession, this precession rate would be roughly $\frac{m_{\rm disk}}{m_\bigodot} n_{\rm N}$, where $n_{\rm N}$ is Neptune's orbital frequency. Thus a disk mass of roughly 60 Earth masses would produce a precession period of 0.9 Myr. Therefore, it is unlikely that the planetesimal disk alone could produce sufficiently fast precession. When Neptune is in the region from 28-29 AU -- and the classical region is sandwiched between the 9:5 and 2:1 resonance -- oscillations in Neptune's semimajor axis due to resonant interactions with Uranus could cause a chaotic sea that extends into this ``sandwich" region. Thus fast precession may fail for parameters for which fast damping has the possibility of working. 

We will present a contour map for Neptune's precession rate, analogous to Figure \ref{fig:ccold}, in Section \ref{sec:combo}.

\subsection{Constraints on transporting the hot objects to the classical region}
\label{sec:hot}

Now we consider which parameters of Neptune will allow the transport of the hot objects from the inner disk into the classical region. We consider the full range of $(a,e)$ into which Neptune can scatter objects from the inner disk and the resulting secular oscillations of the objects once they reach the classical region. We calculate a \emph{minimum} eccentricity damping time as a function of $(\aN, \eN)$ by requiring that particles can reach the stability region defined by $q > 34$ AU (CONSTRAINT 2). The ``minimum time" (contoured) is the time at which it is possible for objects in 50$\%$ of semimajor axes intervals, $\Delta a =$ 0.063 AU, to reach this stable region.

\begin{figure}[htbp]
\begin{centering}
\includegraphics[width=.5\textwidth]{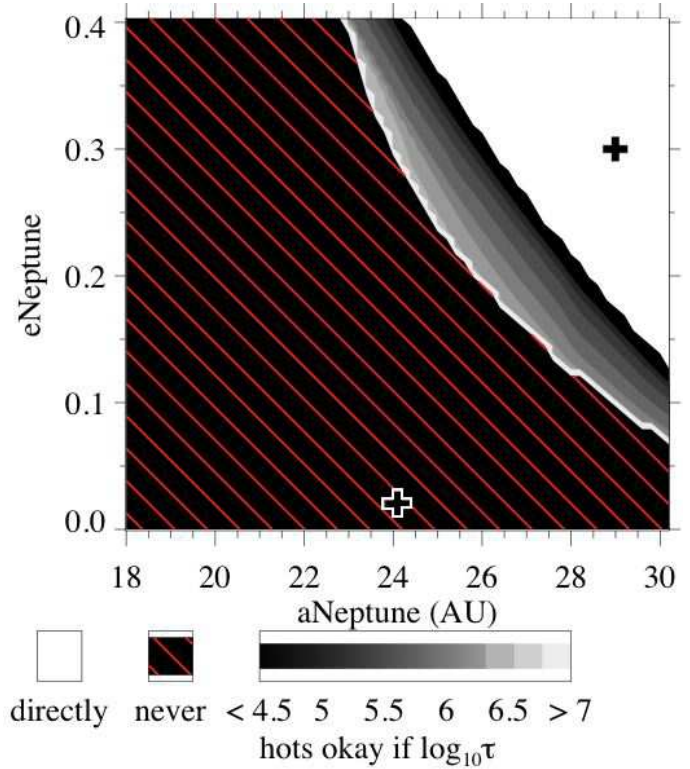}
\caption{Parameter space of Neptune consistent with delivering the hot objects from the inner disk to the classical region, where $a_{\rm Neptune} = \aN$ and $e_{\rm Neptune} =\eN$ are spelled out for clarity. The white region in the upper right is where particles are immediately scattered into the stable region, $q < 34$ AU. The bottom left region (black with red stripes) is where particles can never reach the stable region because their forced eccentricity is too small relative to their free eccentricity. In the contoured region (middle), particles can secularly evolve into the stable region after being scattered if Neptune's eccentricity remains high for long enough. The contours represent the minimum time (see the text) for any scattered KBOs to evolve into the stable region. The pluses mark the conditions of the top panel and the bottom panel of Figure \ref{fig:scatterallow}, which illustrates the region into which Neptune can scatter objects for two sets of illustrative parameters. \label{fig:chot}}
\end{centering}
\end{figure}

Figure \ref{fig:chot}, a map of constraints on Neptune's parameters that deliver the hot objects, illustrates that there are three outcomes for the transport of hot objects:
\begin{enumerate}
\item White region: If $r_{a,{\rm N}}$ is sufficiently large, some objects are scattered into the stable region. This criterion can be quantified as
$ r_{a,{\rm N}} = \aN (1+\eN) > 34 $ AU.
\item Black region with red stripes: If $r_{a,{\rm N}}$ is too small, the KBOs are scattered into a region that is not stable over 4 Gyr and their forced eccentricity is too small (relative to $e_{\rm free}$) to allow them to secularly evolve down into the stable region. This second outcome can be understood from Equation (\ref{eqn:sec}). If the forced eccentricity is small, the free eccentricity will be close to the particle's initial (large) eccentricity after scattering. Thus the free and forced eccentricity vectors cannot destructively cancel to achieve a low total eccentricity.
\item Middle region with gray-scale timescale contours: For intermediate values of $r_{a,{\rm N}}$ , the particles are scattered above the stable region (i.e. to eccentricities too large for stability) but can secularly evolve down into the stable region if Neptune's eccentricity damping timescale is long enough. The time contours refer to the minimum time (defined above) for particles to secularly evolve to the stable region. We note that, because the analytical model is only first order in the KBO's eccentricity, the timescales in the intermediate region may be in error by up to a factor of two (see Appendix \ref{subsec:refined} for discussion). However, the particles most relevant for this scenario --- those rapidly declining in eccentricity --- are matched with substantially smaller error. We also note that this regime (Outcome 3) is a narrow region of parameter space and that most parameters for Neptune fall within one of the two regimes above.
\end{enumerate}

The minimum values for $(\aN, \eN)$ in the intermediate range of $r_{a,{\rm N}}$ has a messy analytical solution. The minimum eccentricity a particle can achieve is:
\begin{equation}
\label{eqn:emin}
e_{\rm min} = (e_{\rm free}^2+e_{\rm forced}^2-2e_{\rm free}e_{\rm forced})^{1/2}
\end{equation}
Combining with Equation (\ref{eqn:scatter0}) and (\ref{eqn:efree}), for a given $\aN$ one can solve for the cutoff $\eN$, below which particles cannot evolve into the long-term stable region.

\subsubsection{Effects of resonances and chaos}
\label{subsec:hotreschaos}

As shown in Section \ref{subsec:coldres}, the secular forcing near resonance is much faster than in regions outside of resonance. Here we address the effects of resonances for the three outcomes we discussed above:
\begin{enumerate}
\item	 If $r_{a,{\rm N}}$ is sufficiently large, some objects are scattered directly into the stable region. CONSTRAINT 2 ($q > 34$ AU) can always be met in this region, regardless of secular evolution, so we consider this region to always work for the hot classicals. Mean motion resonances can make this already-allowed region even better, allowing even more objects to reach the stable region.
\item If $r_{a,{\rm N}}$ is small, objects are scattered to eccentricities that are too large to allow the objects to reach the stable region, even through fast secular evolution near resonance. Although fast secular evolution near resonance lets objects reach their minimum eccentricity (Equation \ref{eqn:emin}), unfortunately all the objects have minimum eccentricities too large for long-term stability.
\item In the intermediate regime (the small contoured strip of parameter space in Figure \ref{fig:chot} that does not deliver the hot classicals ``directly" or ``never"), objects are not scattered directly to the stable region but can reach the stable region through secular evolution, including fast secular evolution near resonances. Without resonances, the secular evolution timescale at 47.5 AU (the largest KBO semimajor axis) sets the minimum timescale over which Neptune needs to maintain its high eccentricity. However, because of fast secular evolution near resonance, the timescale is actually set by the largest KBO semimajor axis unaffected by proximity to resonance. Thus the minimum timescale is shorter (i.e. Neptune can damp more quickly) than if we neglected the effects of resonances. In practice, the secular evolution time is not a steep function of the particle's semimajor axis \citep[Figure 4]{2012W}. We created another version (not shown) of Figure \ref{fig:chot} without including the effects of mean motion resonances. We required objects at 47.5 AU to be able to reach $e < 0.3$, because this would ensure that objects interior of 47.5 would also reach the stable region. The results were qualitatively the same, with some slightly longer minimum times.
\end{enumerate}

Now we consider whether objects that are actually \emph{in} resonance, as opposed to being \emph{near} resonance, could be more easily delivered to the classical region than our constraints would imply. Although we did not explicitly take into account objects actually librating in resonance, we argue that, as a by-product, our constraints include this effect. When Neptune's eccentricity is low or its semimajor axis places no strong resonances in the classical region, objects actually-in-resonance with Neptune occupy a very small portion of phase space. In the opposite situation, in which Neptune's eccentricity is high and strong resonances are present in the classical region, the parameters for Neptune are already in Outcome 1 above (``directly" scattered). For example, the initial conditions used to create the Figure 3 of \citet{2008L} are part of this region. The constraints we will place in this region will come only from the cold objects.

Potentially the 2:1 resonance could ``sweep" the classical region, rapidly evolving transported objects into the stable region, as Neptune migrates. Moreover, resonances sweeping the region could deliver hot objects to the stable region via orbital chaos.  However, neither of these scenarios would be consistent with preserving the cold classicals. Because the hot objects span the whole region from $42.5$ AU $< a < 47.5$ (Figure 3, right panel), the 2:1 resonance would have to sweep all the way through this range, exciting the cold classicals (as would the 9:5 resonance as it passes 42.5 AU). Moreover, if interactions of Neptune and Uranus created a chaotic sea (Section \ref{subsec:chaos}), the cold classicals would be dynamically disrupted. Therefore, the location of resonances do not provide additional constraints or possibilities for the transport of hot classicals.

\subsubsection{Effects of precession}

Precession has several effects on delivery of the hot classicals:

\begin{enumerate}
\item Before the eccentricity of Neptune damps, precession allows Neptune to scatter more objects from the inner disk. If a KBO in the region near Neptune is on an eccentric orbit, the orbit of Neptune and the small body will not necessary cross, depending on the relative positions of their periapses. But if Neptune's periapse quickly precesses, Neptune's orbit may come to intersect additional KBO orbits.
\item As an object secularly evolves after being transported to the classical region, its longitude of periapse varies.  Hence, precession of Neptune's orbit changes the likelihood that the object will be scattered again before it reaches the long-term survival region. We note that the likelihood of scattering depends on the \emph{relative} precession rates of Neptune and the object.
\item If the precession rate is fast enough, it decreases the forced eccentricity. The magnitude of the particle's free eccentricity is given by Equation (\ref{eqn:efree}). For particles with quickly decreasing eccentricity ($\varpi - \varpiN = \pi/2$) (Equation \ref{eqn:edot}), a reduction in $e_{\rm forced}$ decreases the amplitude of the free eccentricity. Thus fewer particles scattered above the stable region are ever able to reach it. Certain $(\aN,\eN)$ that allowed particles to evolve into the stable region, when Neptune was not quickly precessing, will no longer work.
\item The precession of the forced eccentricity allows particles that are evolving down into the stable region via secular oscillation to reach the region more quickly. This effect changes the timescales for objects evolving to the stable region. 
\end{enumerate}

Thus precession can either help or hurt the transport of hot objects into the classical region, depending on the particular combination of parameters. 

\subsubsection{Scattering efficiency of the hot objects}
\label{subsec:efficient}

The number of hot objects that reach the stable region depends on the surface density of planetesimals in the inner disk, the eccentricity damping timescale of Neptune, and Neptune's precession rate.  A faster precession rate evidently delivers more objects to the classical region, but objects are also more likely to exit the classical region via subsequent frequent scatterings, resulting in a steady state flux. Perturbations from other planets could increase the efficiency by increasing Neptune's precession rate and causing additional perturbations to the inner disk. We leave a more detailed exploration of the scattering efficiency for future work. 

However we contrast the inefficient process of transporting the hot objects into the classical region to the highly efficient preservation of the cold objects, i.e. all the cold objects stay in the classical region but few hot objects are transported there. Such a discrepancy might be possible if the inner disk is much denser than the outer disk. A dense inner disk is consistent with the short eccentricity damping timescales we are finding (Section \ref{sec:cold}), because the damping rate scales with the surface density of planetesimals \citep{2007F}. A dense inner disk and rarefied outer disk may explain why Neptune ceased its planetesimal-driven migration when it reached 30 AU \citep[see][for the suggestion that the planetesimal disk was truncated at 30 AU]{2003L}. However, the low number of cold objects poses a problem for their formation, as we will discuss in Section \ref{sec:discuss}.

\subsection{Combined constraints from both hot and cold objects}
\label{sec:combo}

We placed constraints on parameters of Neptune that preserve the confinement of cold objects to low eccentricities (Section \ref{sec:cold}) or allow the transportation of the hot objects from the inner disk into the classical region (Section \ref{sec:hot}),. These constraints are useful separately, and they confirm in situ formation as a feasible origin for the cold objects and transport from the inner disk as a feasible origin for the hot objects. In this Section, we investigate which parameters permit the combination of these two origins, producing both a cold and a hot population. This may be possible if there is overlap between the parameter space that preserves the cold classicals (Section \ref{sec:cold}) and the parameter space that transports the hot classicals (Section \ref{sec:hot}).

\begin{figure}[htbp]
\begin{centering}
\includegraphics[width=.5\textwidth]{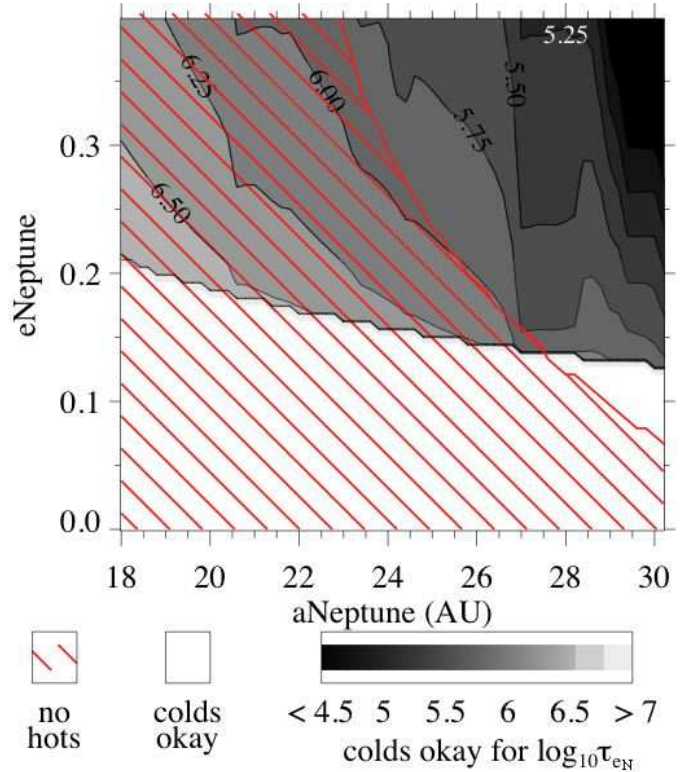}
\caption{Combined constraints from the hot and cold classicals when the cold classicals are preserved by Neptune damping quickly, where $a_{\rm Neptune} = \aN$ and $e_{\rm Neptune} =\eN$ are spelled out for clarity. The contours refer to the maximum eccentricity damping time $\log_{10} \tau_{\eN}$ (where $\tau_{\eN}$ is in years) for Neptune in this region to avoid excessively exciting the cold classicals. In the white region, the forced eccentricity imparted by Neptune in the region from 42.5 to 45 AU is below 0.1. In the red, diagonally striped region, Neptune cannot transport the hot objects to the classical region, defined as $q > 34$ AU from 42 to 47.5 AU. }\label{fig:combo}
\end{centering}
\end{figure}

In Figure \ref{fig:combo}, we combine the constraints from the hot and cold classicals. First, as a function $(\aN, \eN)$ we plot the contours for the maximum eccentricity damping time necessary to keep the cold classicals at $e < 0.1$ in the region from 42.5 to 45 AU. In the white region, Neptune's eccentricity is low enough such that the final eccentricities of the cold objects will be below 0.1, no matter how long the damping timescale. Then we overplot red, diagonal stripes in the region where Neptune cannot deliver the hot classicals, neither by direct scattering nor post-scattering secular evolution. In creating this region, we took into account the sliver of parameter space (Figure \ref{fig:scatterallow}, contoured region) for which Neptune cannot scatter objects directly into the region $q > 34$ AU but for which objects can evolve into the stable region through secular evolution. We compare the time required for the hot objects to reach the long-term stable region to the maximum eccentricity damping time to preserve the cold classicals. In most cases, Neptune's eccentricity must damp before any hot objects reach the long-term stable region. From all these considerations, there is only a  small region of parameter space where Neptune can deliver the hot classicals (no red, diagonal stripes) while fast secular evolution near resonances does not quickly excite the cold classicals (light regions). 

The allowed region in ($\aN, \eN$) space, in which the hot objects can be transported without disrupting the cold population, is bounded by a strip extending from (23 AU, 0.4) to (30 AU, 0.10) on the left and (27.5 AU, 0.4) to (30 AU, 0.2) on the right. To the top right of this strip, the series of resonances extending from the 5:2 to the 9:5, widened by Neptune's high eccentricity, are on top of the cold classical region. Within this bounding strip is a forbidden region near $\aN = 27.5$ AU, at which the 2:1 resonance overlies the cold classical region.

Based on the considerations above, lower limits can be placed on Neptune's migration timescale while Neptune's eccentricity is high. During Neptune's period of high eccentricity, Neptune's migration timescale must be slow enough to keep Neptune in a region that will not quickly excite the cold classicals. Neptune should not spend substantial time with high eccentricity near $\aN = 27.5$ AU or $\aN = 30$ AU, lest resonances disrupt the classical region. Thus Neptune is constrained to migrate no more than a few AU during Neptune's eccentricity damping time. Because of the discrete ranges of consistent semimajor axes, when resonances are accounted for, a ``damp first and then migrate" scenario is consistent with preserving the cold classicals, while a ``migrate first and then damp" scenario is not. 

\begin{figure}[htbp]
\begin{centering}
\includegraphics[width=.5\textwidth]{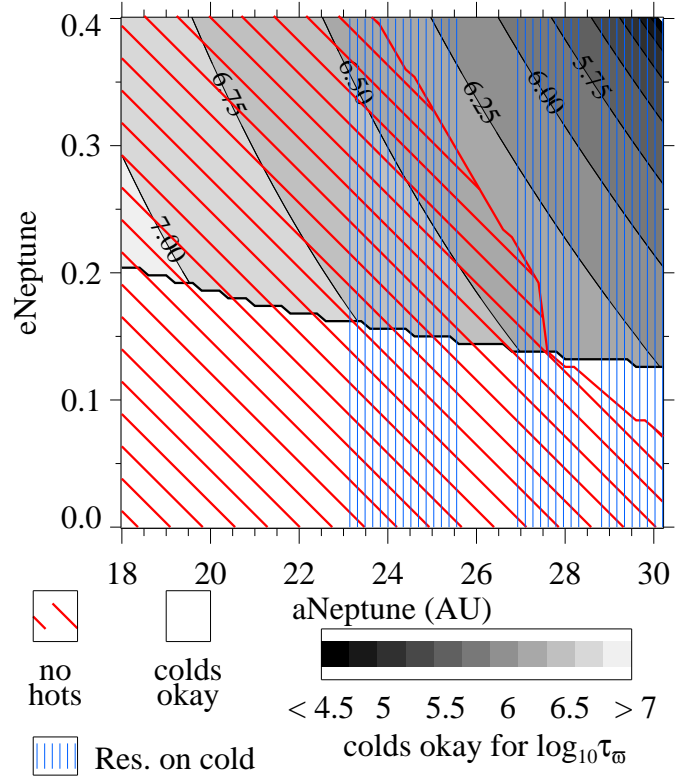}
\caption{ Combined constraints from the hot and cold classicals when the cold classicals are preserved by Neptune precessing quickly, where $a_{\rm Neptune} = \aN$ and $e_{\rm Neptune} =\eN$ are spelled out for clarity. The contours refer to the maximum precession period $\tau_\varpi = 2\pi/\dot{\varpi}_{\rm N}$ for Neptune to avoid excessively exciting the cold classicals. Neptune must precess fast enough to keep $e_{\rm forced} < 0.1$ and, when Neptune intrudes into the region, $e_{\rm forced} < \frac{1}{2} (1-q_{a,{\rm N}}/42.5)$. In the white region, the forced eccentricity imparted by Neptune in the region from 42.5 to 45 AU is below 0.1 (and below $\frac{1}{2} (1-q_{a,{\rm N}}/42.5$), even when Neptune does not precess). In the red, diagonally-striped region, Neptune cannot transport the hot objects into the stable classical region, defined as $q > 34$ AU from 42.5 to 47.5 AU. The blue, vertically striped regions denote $\aN$ for which the center of a resonance, of fourth order or lower, lies in the region from 42.5 to 45 AU. These regions are not necessary forbidden (see the text).}\label{fig:combopre}
\end{centering}
\end{figure}

In Figure \ref{fig:combopre}, we plot analogous constraints for the special case of fast precession. The contours represent the precession rate of Neptune necessary to keep $e_{\rm forced}$ sufficiently low (Equation \ref{eqn:extraforced}). In all cases, the KBOs' forced eccentricities must be below $e_{\rm forced} < 0.1$ (CONSTRAINT 1). We impose an additional constraint to avoid the scattering of cold particles as they reach their maximum eccentricities of $e_{\rm max} = 2 e_{\rm forced}$. We require $e_{\rm max}$ must be below the scattering line at the inner edge of the cold classical region: therefore $e_{\rm forced} < \frac{1}{2} (1-q_{a,{\rm N}}/42.5)$. For large values of Neptune's apoapse $q_{a,{\rm N}}$, this constraint is stricter than $e_{\rm forced} < 0.1$. This constraint may, in fact, be too strict because if a cold classical KBO is scattered by Neptune, it is more likely to scattered out of the classical region altogether than to end up between $42.5$ AU $< a < 45$ AU at an eccentricity too large to be consistent with the observations but small enough to survive over 4 Gyr. We leave a detailed investigation into the role of scattering in the case of fast precession for future work. The red, diagonally striped region is where hot objects cannot reach the long-term stable region. 

Next we consider the constraints for delivering the hot classicals. For each $(\aN,\eN)$, we calculate whether objects can be directly scattered into the region $q > 34$ AU. If not, we assume Neptune imparts a forced eccentricity small enough to be consistent with the constraints above for keeping the cold classicals unexcited (i.e., $e_{\rm forced} = 0.1$, or, if smaller, $e_{\rm forced} < \frac{1}{2} (1-q_{a,{\rm N}}/42.5)$) and determine whether the hot object can reach the stable region through secular evolution, using Equation (\ref{eqn:emin}). If Neptune cannot directly scatter objects into the stable region and if objects secularly evolving cannot reach $q > 34$ AU in more than $50\%$ of semimajor axis intervals, $(\aN, \eN)$ is not consistent with delivering the hot objects. In the cases where fast precession is necessary to keep the cold classicals confined to low eccentricities, hot classicals scattered into the region have the same low forced eccentricities; therefore, they do not experience significant secular evolution down into the classical region but remain at their post-scattering eccentricities. 

Finally, in blue vertical stripes, we overplot ranges of $\aN$ for which the centers of one or more resonances of fourth order or lower lie in the cold classical region from 42.5 to 45 AU. These $\aN$ are not necessary forbidden, but chaos can occur if Neptune's eccentricity is high and interactions with Uranus cause oscillations in $\aN$.

Though we are applying constraints that are lenient and conservative in ruling out regions of parameter space, a substantial fraction of parameter space is ruled out. The following scenarios have not been ruled out and may work, depending on the details of the interactions between Uranus and Neptune. In both of these scenarios, resonances cannot overlie the region of $42.5$ AU $< a < 45$ AU.

\begin{itemize}
\item Short $\tau_e$ and large $r_{p,N}$: An eccentric Neptune transports objects into the classical region until its eccentricity damps, which occurs quickly enough that the cold classicals are not excited (Figure \ref{fig:combo}).
\item Fast $\dotvarpiN$ and large $r_{p,N}$: An eccentric, quickly precessing Neptune transports objects into the classical region and its quick apsidal precession keeps the forced eccentricity of the particles low, preserving the cold classicals (Figure \ref{fig:combopre}).
\end{itemize}

In principle, an intermediate $r_{p,N}$ could deliver hot objects to high eccentricities $(e > 0.3)$ in the classical region but allow them to evolve to lower eccentricities $(e < 0.3)$ before the cold objects are excited. In practice, we found the timescales are not compatible: if the eccentricity damps quickly enough to preserve the cold classicals from secular excitation, there is not time for hot objects to secularly evolve down into the Belt.  However, for parameters of Neptune for which hot objects are scattered directly in the stable region, secular evolution can allow the objects to reach even lower eccentricities, especially the objects undergoing fast secular evolution near resonances.

Therefore, in practice, the consistent dynamical histories are ones in which Neptune has a large enough apoapse to transport hot objects immediately into the stable region, with its eccentricity damping quickly enough or precessing quickly enough so that the cold classicals remain at low eccentricities consistent with the observations.

\subsection{Interpretation of constraints in light of Neptune's full dynamical history}
\label{subsec:combointerpret}

The goal of this paper is to determine which parameters for Neptune allow the planet to deliver the hot classicals from the inner disk into the classical region without dynamically disrupting the in situ cold classicals. Throughout its dynamical history, Neptune must satisfy the constraints presented here to avoid excessively exciting the cold classicals (Section \ref{sec:cold}). At some point, Neptune must also spend time in a region of parameter space where it can also deliver the hot classicals (Section \ref{sec:combo}). We clarify that Neptune did not necessary form at the location where hot classical delivery takes place or arrive there after undergoing a single, instantaneous scattering. Before and after hot classical delivery, Neptune can potentially spend time in any region of parameter space as long it obeys our constraints against not excessively exciting the cold classicals. The constraints developed here, which can serve as a ``road map" for Neptune's path through parameter space, hold for realistic models that include multiple scatterings and have a straightforward interpretation.

In the context of the Nice model, Neptune may have undergone a series of scatterings, spending time at a variety of spots in $(\aN, \eN)$ space. In each of these spots, Neptune must obey the constraints we place to avoid disrupting the cold classicals. Perhaps the scattering occurs quickly compared to the excitation time for the cold classicals. If not, Neptune could pre-excite the cold classicals (but only to below the observational limit) before it reaches a region where it can deliver the hot classicals. In imposing our conservative constraints, we are assuming that the cold classicals begin with $e=0$, but if they are pre-excited, the constraints will be stricter.

Another possibility is that Neptune may spend a long time at a location where it creates a very small region of stability in the classical region, clearing out most of the cold classicals. This is a potential solution to the mass efficiency problem, which we will discuss in Section \ref{sec:discuss}.  However, as shown in Section \ref{subsec:notstability}, Neptune cannot deliver the hot classicals in this regime. Therefore, after this period has ended, when Neptune is delivering the hot classicals, the constraints we will place on not exciting the cold classicals will hold.

The scattering(s) Neptune undergoes are quick changes in its orbit. After its period of high eccentricity ends -- or during a temporary period of low eccentricity -- Neptune can undergo slow evolution, including slow migration and slow damping of its (now small) eccentricity. The KBOs will maintain their free eccentricities throughout this slow evolution. If Neptune's eccentricity were excited very gradually, on a timescale much longer than the secular evolution times of the cold classicals, the cold classicals would keep their initial low free eccentricities. However, we expect the excitation of Neptune's eccentricity via scatterings to take place on a timescale shorter than millions of years \citep[e.g.][]{1999T}.

\begin{figure}[htbp]
\begin{centering}
\includegraphics[width=.5\textwidth]{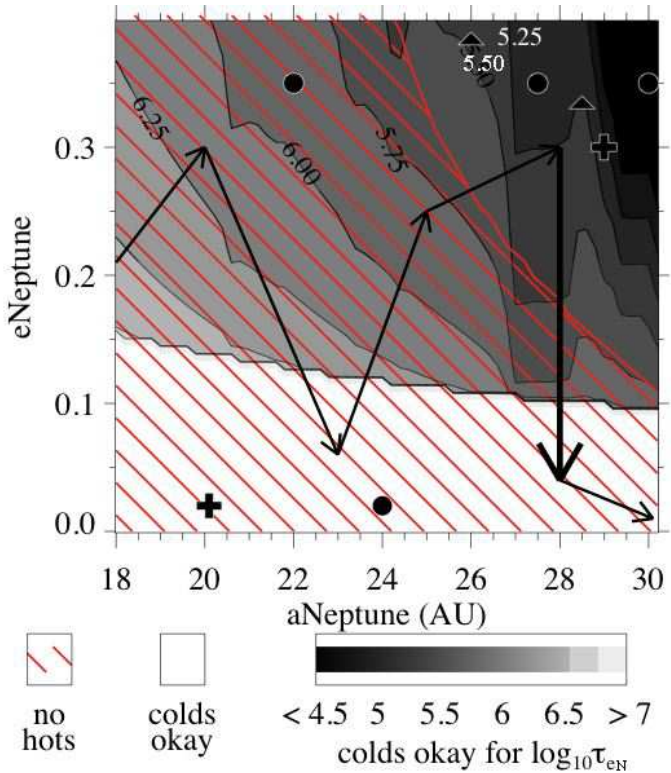}
\caption{ Combined constraints from the hot and cold classicals, where $a_{\rm Neptune} = \aN$ and $e_{\rm Neptune} =\eN$ are spelled out for clarity. The contours refer to the maximum eccentricity damping time $\log_{10} \tau_{\eN}$ (where $\tau_{\eN}$ is in years) for Neptune in this region to avoid excessively exciting the cold classicals. The constraints are stricter than in Figure \ref{fig:combo}. In the white region, the forced eccentricity imparted by Neptune in the region from 42.5 to 45 AU is below 0.075. In the red, diagonal striped region, Neptune cannot transport the hot objects to the classical region, defined as $q > 36$ from 42.5 to 47.5 AU. The arrows are a schematic illustration of an example of Neptune's path through parameter space as it undergoes multiple scatterings, having its eccentricity damped, or being re-scattered, on the contoured timescale to avoid excessively exciting the cold classicals. The pluses mark the initial conditions from \citet{2005H} (bottom) and \citet{2008L} (top). The circles are the parameters for example integrations shown in Figure \ref{fig:a26} and the triangles for example integrations shown in Figure \ref{fig:a26goods}.}\label{fig:combopath}
\end{centering}
\end{figure}

We have ruled out much of parameter space with the constraints that Neptune cannot excite the cold objects above $e=0.1$ and must be able to deliver at least a few hot objects to $q > 34$ AU in the region from 42-47.5 AU. We note that we are ``ruling out" parts of parameter space where Neptune cannot deliver the hot objects without disrupting the cold, not ``ruling out" that the planet can ever spend time there (see above). In Figure \ref{fig:combopath}, we look for greater consistency with the observations (Figure \ref{fig:diagnosticall}): a forced eccentricity less than 0.075 for the cold objects and $q_N > 36$ AU (meaning that a hot object could be scattered to an eccentricity as low as 0.24 at 47.5 AU). The parameter space shrinks where Neptune can deliver the hot objects without disrupting the cold (i.e. light gray regions with no red, diagonal constraints). Over-plotted on this figure is an example (arrows) of Neptune's path through parameter space. In this example, Neptune first undergoes multiple scatterings, spending a short enough time at each $(\aN,\eN)$ to avoid exciting the cold classicals. Then it reaches 28 AU with an eccentricity of 0.3; here it delivers the hot classicals, as its eccentricity damps quickly enough to avoid exciting the cold classicals. Then it migrates at a low-eccentricity to its current location.

On this figure, we overplot some parameters as symbols. For the triangles and circles, we will show example integrations in Section \ref{subsec:example}. The pluses are parameters taken from the literature. The plus at $\aN = 20, \eN = 0.02$ marks the initial condition of \citet{2005H}, from which Neptune undergoes migration to its current location. Neptune remains in a region that is white (never excites the cold classicals) but also red, diagonal striped (does not deliver the hot classicals); thus this simulation maintained the low eccentricities of the cold classicals but did not deliver the hot classicals. The plus at $\aN = 28.9, \eN = 0.3$ indicates the initial condition for \citet{2008L}, Run B. In this part of parameter space, Neptune can deliver the hot classicals, which were indeed produced by this simulation. However, Neptune's eccentricity damped on a timescale of $\tau_e = $ 2 Myr, too slow to avoid exciting an in situ cold population.

\subsection{Example integrations illustrating constraints}
\label{subsec:example}

\begin{figure}[htbp]
\begin{centering}
\includegraphics{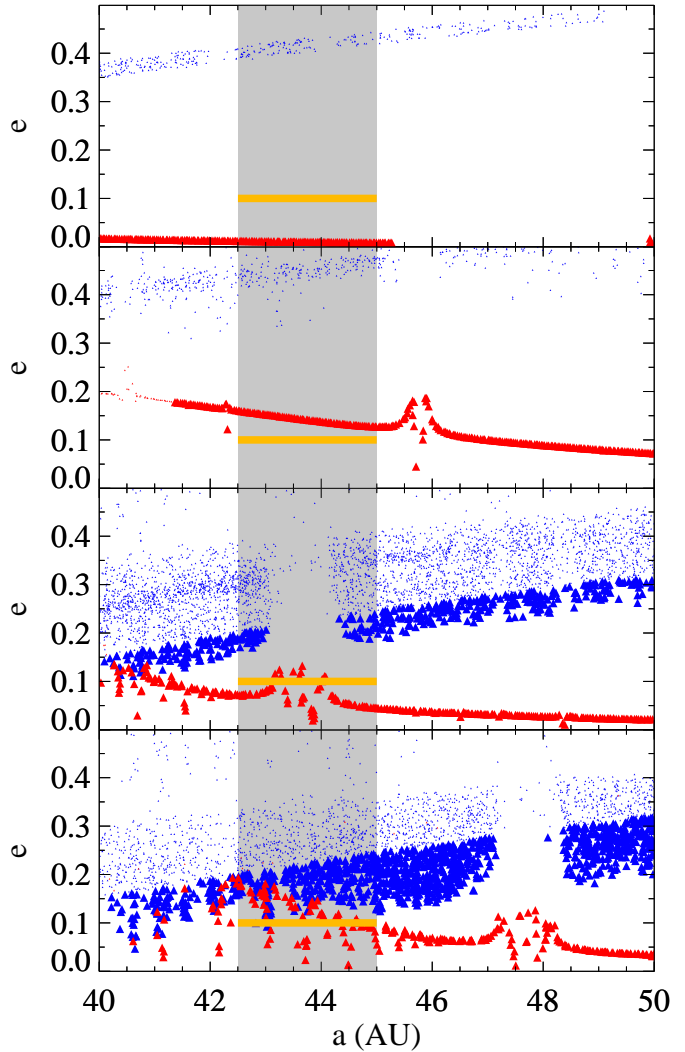}
\caption{ Examples of four numerical integrations that violate the constraints established (Figure \ref{fig:combo}). The blue and red triangles are the positions of particles at the end of the integration. The red triangles are cold objects which began in the region from 40 to 50 AU with $e=0$. The blue triangles are hot objects which began in the inner disk interior to 38 AU. The small triangles have eccentricities above the region of long-term stability (dashed line in Figure \ref{fig:diagnosticall}) and thus are not expected to survive over 4 Gyr. The yellow line indicates $e=0.1$ and the shaded gray region is 42.5 AU $< a < $45 AU, where the cold classicals are observed (Figure \ref{fig:diagnosticall}) to be confined to low eccentricities. The parameters for Neptune in each integration are: $\aN = 24, \eN = 0.02, $no eccentricity damping (top row), $\aN = 22, \eN = 0.35, \tau_e = $2 Myr (row 2), $\aN = 27.5, \eN = 0.35, \tau_e = 0.2$ Myr (row 3), $\aN = 30, \eN = 0.35, \tau_e = 0.2$ Myr (row 4). The snapshots are at times 1.6 Myr, 6 Myr, 1.6 Myr, and 1.6 Myr.} \label{fig:a26}
\end{centering}
\end{figure}

\begin{figure}[htbp]
\begin{centering}
\includegraphics[width=.5\textwidth]{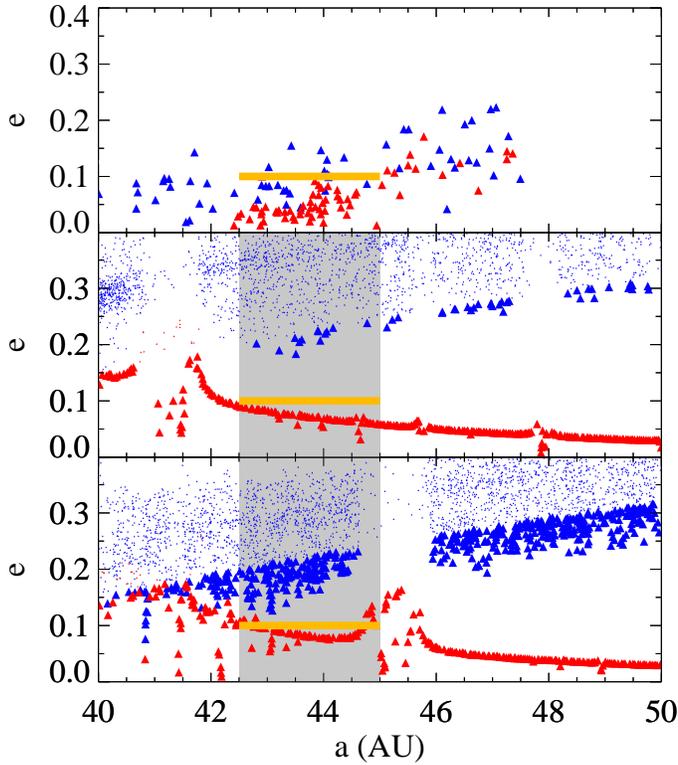}
\caption{ Examples of two integrations that obey the constraints established (Figure \ref{fig:combo}). The top row shows the observed objects from Figure \ref{fig:diagnosticall} for comparison.  In rows 2 and 3, blue and red triangle are the positions of particles at the end of the integration. The red triangles are cold objects which began in the region from 40 to 50 AU with $e=0$. The blue triangles are hot objects which began in the inner disk interior to 38 AU. The small triangles have eccentricities above the region of long-term stability (dashed line in Figure \ref{fig:diagnosticall}) and thus are not expected to survive over 4 Gyr. The yellow line indicates $e=0.1$ and the shaded gray region is 42.5 AU $< a < $45 AU, where the cold classicals are observed (Figure \ref{fig:diagnosticall}) to be confined to low eccentricities. The parameters for Neptune in each integration are: $\aN = 26, \eN = 0.38, \tau_e = $ 0.33 Myr (second row) and $\aN = 28.5, \eN = 0.33, \tau_e = $ 0.25 Myr (third row). \label{fig:a26goods}}
\end{centering}
\end{figure}

Example integrations illustrating the constraints we have derived are shown in Figure \ref{fig:a26} and \ref{fig:a26goods}. Each integration lasts for 1.6 Myr unless otherwise noted (see Section \ref{subsec:short} for a general description of the integrations). In addition to the 600 test particles in the region from 40-60 AU, we add 59600 test particles in the region from 18-38 AU, each with initial $e = i = 0$, representing the inner planetesimal disk from which the hot classicals are scattered.

The parameters for the four integrations shown in Figure \ref{fig:a26} are plotted in the parameter space map in Figure \ref{fig:combopath} as circles and correspond to regions of parameter space where Neptune cannot both deliver the hot objects and keep the cold objects at low eccentricities. In row 1, Neptune is at $\aN = 24, \eN = 0.02$ and undergoes no eccentricity damping. The cold classicals remain confined to low eccentricities, but the hot classicals are delivered to high eccentricities and cannot secularly evolve down to lower eccentricities. For the same reason (i.e. that Neptune's apoapse is too small), the hot classicals are also not delivered in row 2 ($\aN = 22, \eN = 0.35$). The damping timescale in this integration is longer (2 Myr) than half a secular evolution time, so the eccentricities of the cold objects converge to the forced eccentricities, which are above 0.1. This integration lasts 6 Myr. In the third row ($\aN = 27.5, \eN = 0.35, \tau_e = 0.2$ Myr) and fourth row ($\aN = 30.06, \eN = 0.35, \tau_e = 0.2$ Myr), the hot classicals are delivered but the cold classicals are excited by fast secular evolution near resonances.

The parameters for the two integrations shown in Figure \ref{fig:a26goods} are plotted in the parameter space map in Figure \ref{fig:combopath} as triangles and correspond to a set of parameters in each of the two viable regions of parameter space. In both cases, Neptune can deliver the hot classicals. Moreover, since it is in a region of parameter space where no resonances overlie the region of 42.5 AU $< a <$ 45 AU, and since its eccentricity damps quickly enough to obey our constraints, the cold classicals remain at $e < 0.1$. In the top panel, we plot the observed objects from Figure 3 for comparison. In the middle panel, the 2:1 resonance is interior to the cold classical region. In the bottom panel, the cold classicals are sandwiched between two regions where the cold classicals are excited by fast secular evolution near resonance. 

\section{Discussion}
\label{sec:discuss}
Through an exploration of parameter space (Section \ref{sec:constraints}) combined with conservative criteria from the observed eccentricity distributions of classical KBOs (Section \ref{subsec:features}), we reach the conclusion that most parameters for Neptune are inconsistent with both delivering a hot population from the inner disk and preserving a cold population formed in the outer disk. We have explored the full parameter space for a generalized model --- in which Neptune undergoes some combination of high eccentricity, migration, and/or precession and delivers the hot objects on top of the cold --- that encompasses the previous ``chaotic capture" and ``extensive migration" models and accounts for the different physical properties of the hot and cold classicals. We have found that the generalized model remains viable only in two restricted regions of parameter space: Neptune is scattered onto an eccentric orbit with a semimajor axis in one of two ranges, 24-27 AU (so that the 2:1 resonance is interior to 42.5 AU) or 28-29 AU (so that the region from 42.5 to 45 AU is sandwiched between the 9:5 and the 2:1 resonance). Although $\aN = 30$ AU appears feasible on the parameter space plots for $\eN \sim 0.15$, in this region ``strips" are excited by overlying resonances and the hot objects are far from low enough. Neptune scatters the hot objects from the inner disk into the stable classical region where we observe them. Because Neptune's eccentricity damps or the planet's orbit precesses quickly, Neptune does this without exciting the cold objects above their observed eccentricities. Because Neptune is confined to one of these two particular regions, mean motion resonances --- which would quickly excite the cold classicals through accelerated secular forcing and/or chaos --- do not, while Neptune's eccentricity is high, overlie the region where we observe the cold classicals confined at low eccentricities today. Most likely, once Neptune's eccentricity damps, it migrates on a circular orbit to its current location, a migration distance of $\Delta \aN$ = 1-6 AU. 

Our constraints should be interpreted in light of Neptune's full dynamical history, which may include multiple scatterings of Neptune and/or excitation/sculpting of the cold classical region before Neptune delivers the hot classicals. Throughout its path through parameter space, Neptune must obey our constraints on not excessively exciting the cold classicals. Whatever the prior early evolution, Neptune must eventually spend time in a region of parameter space where it can deliver the hot classicals while its eccentricity damps or precesses quickly enough --- or Neptune is re-scattered quickly enough-- to avoid exciting the cold classicals. Then it can proceed to its current location via additional scatterings or migration, maintaining an eccentricity low enough to continue to avoid exciting the cold objects.

The viable regions of parameter space are qualitatively and quantitatively different from the previous models that did not produce the observed eccentricity distributions. Compared to the \citet{1995M,2005H} model, Neptune undergoes a period of high eccentricity and migrates a shorter distance ($< 6$ AU, as opposed to 7-10 AU). The most significant differences from the \citet{2008L} model are that the cold population forms in situ, that a fast damping (0.4 Myr as opposed to 3 Myr) or precession rate is required, and that resonances, rather than being the mechanism for creating the cold population by overlapping to create a chaotic sea, cannot overlie the cold classical region while Neptune's eccentricity is high.

Another key finding is that the ``chaotic sea" that may have existed in the classical region during Neptune's wild days \citep{2008L} would not have been caused solely by Neptune's high eccentricity but by oscillations in Neptune's semimajor axis due to its near-resonant interactions with Uranus. Thus the exact dynamical configuration of Uranus and Neptune controls the extent and existence of a chaotic zone. Configurations of these two planets in which their interactions are especially strong might rule out the region of parameter space of $28$ AU $< \aN < 29$ AU during Neptune's high eccentricity period, which corresponds to the classical objects being sandwiched between the 9:5 and the 2:1. A detailed investigation of the effects of the interactions between Uranus and Neptune will likely provide additional constraints on the dynamical history of the solar system. The conditions for the chaotic sea not to disrupt the cold classicals may rule out additional parameter space but will not open up more.

There is a major outstanding problem with forming the cold classicals in in situ: unsettlingly, the current surface density of cold classicals is thought to be too low for the in situ formation of the $~100-1000$ km objects we observe \citep{1997S,1998K}. One potential explanation is that the population has lost substantial amount of mass to collisions and subsequent removal by radiation forces. Another potential resolution is that, given that the physics of planetesimal formation is currently poorly understood, it may be possible to form such large objects at such low surface densities (if some major physical process is missing from our understanding of planetesimal formation). Finally, as discussed in Section \ref{subsec:combointerpret}, the cold classical region may have been depleted by scattering before the period of hot classical delivery. Though none of these potential solutions have been validated, in situ formation of the cold classicals remains viable due to their distinct physical properties.

Our results are intended to provide constraints for extensive numerical integrations that include all the giant planets, have tens of thousands of test particles, and last for the age of the solar system. We established conservative criteria in order to confidently rule out regions of parameter space; the remaining regions are potentially viable but may be ruled out by additional constraints, including those that depend on the details of the configuration of giant planets. We have focused on the classical KBOs in this paper and have not tried explicitly to match the distribution of resonant KBOs. We expect resonant objects to be produced, within the parameter space we constrained, by a combination of migration \citep{1995M}, chaotic capture \citep{2008L}, and, a new mechanism identified here as being important, fast secular evolution to low eccentricities of hot objects delivered near resonance followed by capture. Detailed matching of the resonant population is beyond the scope of this paper but will likely tighten our constraints.

The scattered disk population may also provide additional constraints on Neptune's dynamical history. Within the generalized model we have explored, these objects originate in the inner disk and are scattered out to beyond 48 AU, or within 48 AU at higher eccentricities and inclinations than the classical objects. If the scattered and hot objects have the same origin, any model must correctly produce their relative number. The scattered disk also contains a number of objects beyond 50 AU occupying high-order mean motion resonances. \citet{2007L} found that in order to produce these resonant objects, Neptune must undergo migration after the Kuiper Belt has been pre-excited out to 50 AU. Our constraints may be consistent with this requirement, since we find that Neptune's eccentricity should damp before it migrates to its current location and that, in one allowed region, unexcited objects are sandwiched between objects excited by fast secular evolution, extending out to the edge of the 2:1 resonance. We leave detailed explorations of constraints from the population of resonant objects in the scattered disk for future work.

The fast damping of Neptune's eccentricity would imply frequent planetesimal scatterings and thus a high surface density of planetesimals in the vicinity of its orbit. Fast precession would imply strong interactions with the other giant planets (or a high surface density disk). Either way, the dynamical histories of Neptune that produce the hot and cold KBOs are very different from the peaceful disk formation that was the paradigm until 15 years ago. The most viable regions of parameter space we identified imply that Neptune underwent a period of high eccentricity but, mercifully for the spared planetesimals that are today's cold classicals, Neptune's ``wild days" were over soon.

\acknowledgments RID gratefully acknowledges support by the National Science Foundation Graduate Research Fellowship under grants DGE 064449, DGE 0946799, and DGE 1144152. The numerical integrations in this paper were run on the Odyssey cluster supported by the FAS Sciences Division Research Computing Group. We thank Gurtina Besla, David Charbonneau, Matija Cuk, Matthew Holman, Kaitlin Kratter, David Latham, Renu Malhotra, Diego Munoz, Darin Raggozine, Schuyler Wolff, and Kathryn Volk for helpful discussions. We thank Konstantin Batygin, Daniel Fabrycky, Darin Raggozine, Kathryn Volk, and Jack Wisdom for insightful comments on a manuscript draft. We are grateful to Hal Levison for many helpful comments on improving this paper. We gratefully acknowledge an anonymous referee for constructive feedback.

\begin{appendix}

\section{Statistical significance of the hot and cold classical eccentricity distributions}
\label{app:stats}

In addition to the qualitative assessment performed in Section \ref{sec:obs}, we also conduct statistical tests of the significance of the confinement of the cold population. We perform the one-dimensional Kolmogorov-Smirnov (KS) test with the null hypothesis that the observed distribution is consistent with being drawn uniformly in $a$ and $e$ and then filtered by the survival map \citep{2005L}. We perform the test separately in two regions, $42 < a < 44$ AU and $44 < a < 45$ AU, because the survival map is different in these two regions. Note that we use 42 AU instead of 42.5 AU in order to increase the sample size. We created the survival map distribution from an initial distribution uniformly spaced in $e < 0.3$ with the ``stability map filter" method described in Section \ref{subsec:complications} for three assumed survival rates: bottom of the range, middle of the range, and top of the range. We increased the number of objects in the survival map distribution until the results converged, which meant we had to use 10,000 objects post stability map filtering. First, for the observed distribution, we compared to the observed ``likely cold" objects with $i < 2^\circ$. The resulting probabilities that the observed distribution is consistent with a population shaped only by long-term stability are summarized in Table \ref{tab:alllt2}.

\begin{table}[hbpt]
\caption{Probability from KS test comparing observed cold objects with $i<2^\circ$ from Minor Planet Center \citep{2008G,2011V} to ``survival map" distribution. ``Low," "mid," and "high" refer to the bottom, middle, and top of the 10$\%$ survival range used by \citet{2005L}. For example, for the survival range of $50-60\%$, low, middle, and high would indicate that $50\%, 55\%,$ and $60\%$ of particles in that $(a,e)$ cell survive. See Section \ref{subsec:complications} and Appendix \ref{app:stats} for details. \label{tab:alllt2}}
\begin{tabular}{llllll}
				&sample size	&low					&mid					&high\\
$42$ AU $< a < 44$ AU	&38			&$10^{-7} $			&$10^{-8}$			&$10^{-8}$\\
$44$ AU $ < a < 45$ AU	&13 			&$0.004 $				&$0.002$				&$0.001$	\\
\end{tabular}

\end{table}

Then we created five alternative samples of cold objects. Instead of choosing objects with $i<2^\circ$ for inclusion in the cold sample, we selected a uniform random number between 0 and 1 for each observed object. If the number was less than the probability that the object is cold (based on the distribution of \citet{2010G}), we included it in the sample. Using the \citet{2001B} distribution of inclinations instead of the \citet{2010G} did not significantly affect our results. The resulting probabilities are summarized in Table \ref{tab:all}.

\begin{table}[hbpt]
\caption{Probability from KS test comparing probabilitistically-selected observed cold objects to ``survival map" distribution.\label{tab:all}}
\begin{tabular}{llllll}
				&sample size	&low					&mid					&high\\
$42$ AU $ < a < 44$ AU	\\
				&60			&$10^{-8} $		&$10^{-10}$		&$10^{-11}$\\
				&64			&$10^{-10} $		&$10^{-11}$		&$10^{-13}$\\
				&69			&$10^{-10} $		&$10^{-12}$		&$10^{-13}$\\
				&67			&$10^{-9} $		&$10^{-11}$		&$10^{-12}$\\
				&69			&$10^{-11} $		&$10^{-12}$		&$10^{-14}$\\												
$44$ AU $ < a < 45$ AU	
				&31 			&$10^{-5} $		&$10^{-5}$		&$10^{-6}$	\\
				&33 			&$10^{-5} $		&$10^{-5}$		&$10^{-6}$	\\
				&30 			&$10^{-4} $		&$10^{-5}$		&$10^{-6}$	\\
				&32 			&$10^{-4} $		&$10^{-5}$		&$10^{-5}$	\\
				&29 			&$10^{-5} $		&$10^{-6}$		&$10^{-6}$	\\								
\end{tabular}
\end{table}

Based on these results, the orbital distribution of observed objects is not consistent with our null hypothesis. The confinement of cold objects to low eccentricities is formally statistically significant.

Then we repeated the tests using only the CFEPS objects. Instead of using the survival map distribution, we took the survival map distribution and applied the CFEPS Survey Simulator. We assumed either \emph{H} magnitudes uniformly distributed between 6 and 8 or randomly selected from the observed classical objects, and the results were insensitive to this choice. First we used the ``likely cold" objects with $i < 2^\circ$. The resulting probabilities are summarized in Table \ref{tab:cfepslt2}.

\begin{table}[hbpt]
\caption{Probability from KS test comparing observed CFEPS cold objects to survey-simulated ``survival map" distribution.\label{tab:cfepslt2}}
\begin{tabular}{llllll}
				&sample size	&low					&mid					&high\\
$42$ AU $ < a < 44$ AU	&12			&$0.13 $				&$0.06$				&$0.04$\\
$44$ AU $ < a < 45$ AU	&11 			&$0.09$				&$0.06$				&$0.03$	\\
\end{tabular}
\end{table}

Then we created five alternative samples of cold objects, as described above for Table 2. The resulting probabilities are summarized in Table \ref{tab:cfeps}.

\begin{table}[hbpt]
\caption{Probability from KS test comparing observed CFEPS cold objects to survey-simulated ``survival map" distribution.\label{tab:cfeps}}
\begin{tabular}{llllll}
				&sample size	&low			&mid			&high\\
$42$ AU $ < a < 44$ AU	\\
				&20			&$0.01$		&$0.003$		&$0.002$\\
				&25			&$0.001$		&$0.0004$	&$0.0002$\\
				&20			&$0.01$		&$0.003$		&$0.002$\\	
				&20			&$0.01$		&$0.003$		&$0.002$\\			
				&22			&$0.001$		&$0.0003$	&$0.0001$\\											
$44$ AU $ < a < 45$ AU	
				&16 			&$0.03 $		&$0.01$		&$0.005$	\\
				&20 			&$0.009$		&$0.004$		&$0.001$	\\
				&20			&$0.009$		&$0.004$		&$0.001$	\\
				&19			&$0.01$		&$0.005$		&$0.002$	\\
				&18			&$0.02$		&$0.007$		&$0.003$	\\				
\end{tabular}
\end{table}

These probabilities using the small sample of CFEPS objects are low, supporting our conclusion that the cold objects are confined to low eccentricities, but this result is statistically marginal. However, we would not expect the observed objects in the full MPC sample to preferentially have lower eccentricities (indeed, in Figure \ref{fig:diagnostic}, the Survey-Simulated survival map distribution follows the survival map closely), so, given our results for the full MPC sample, we expect the significance to increase as the CFEPS sample size becomes larger. 

An alternative statistical test is the Anderson-Darling test, which is more sensitive to the tail of the distribution. However, even though we took measures to avoid contamination, there are likely to be ``contaminating" objects in the observed cold objects that are actually hot. Therefore we do not necessarily want to give the outliers higher weight, so we judge that the KS test is more robust for this purpose. Using the Anderson-Darling test, we obtained similar results in the cases with large sample sizes and somewhat higher probabilities in the low sample size cases. 

\subsection{Proper Elements}

\begin{figure*}[htbp]
\begin{center}
\includegraphics[width=\textwidth]{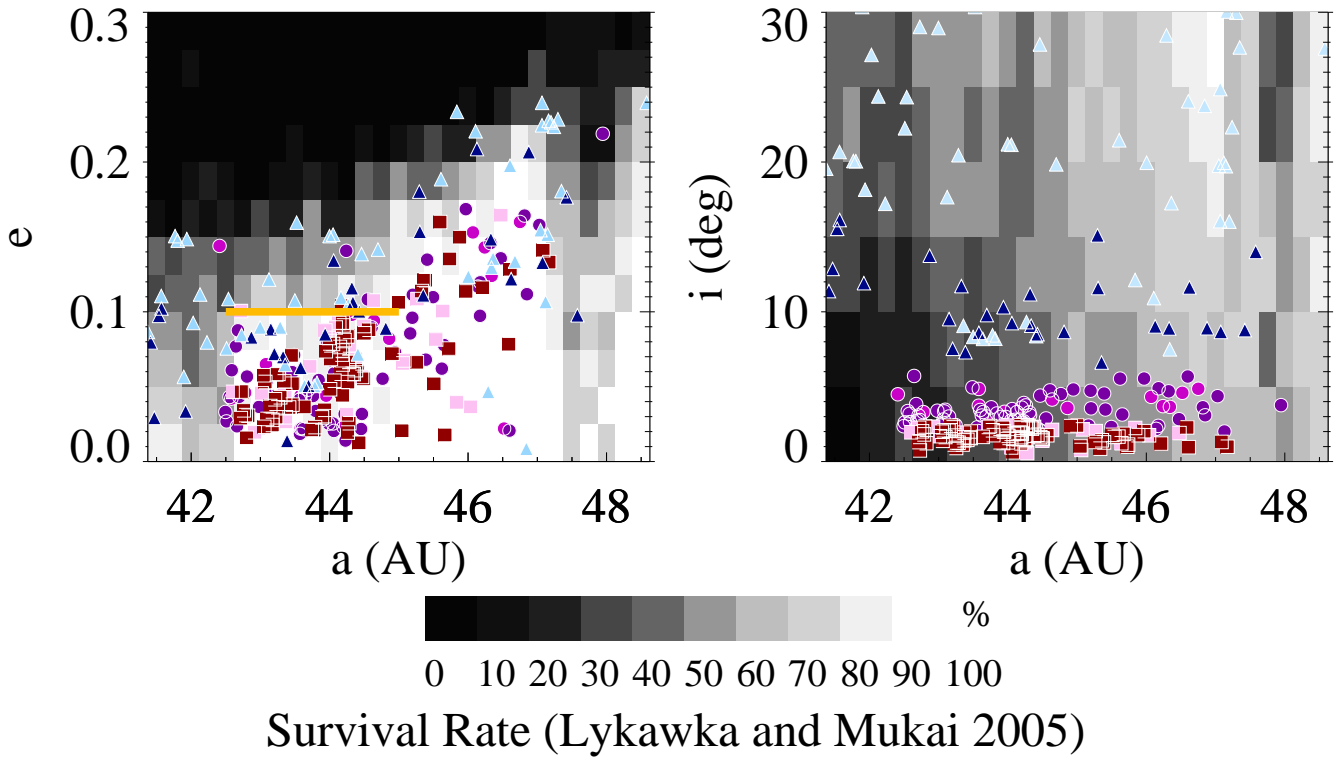}
\caption{Plotted over the survival maps of \citet{2005L} are the proper eccentricity (left) and proper inclination (right) distributions of the observed classical KBOs. Note that the survival map uses instantaneous orbital elements. The red squares are objects with $i < 2^\circ$ and are thus very likely cold classicals. The blue triangles have $i > 6^\circ$ and are thus very likely hot classicals. The membership of any given purple circle ($2^\circ < i <  6^\circ$) is ambiguous. These inclinations are now defined using the proper elements. The light red squares, light blue triangles, and light purple circles, respectively, are objects for which proper elements have not been computed, and thus we plot their instantaneous elements. Classical objects are taken from the Minor Planet Center Database and classified by \citet{2008G} and \citet{2011V}. Proper elements were computed by \citet{2000K,2002K,2003K}. The yellow line indicates our conservative criterion for preserving the cold classicals. \label{fig:proper}}
\end{center}
\end{figure*}

We now consider the free, or proper, elements of the observed KBOs, which have been computed for a subset of KBOs by \citet{2000K,2002K,2003K}. The free elements precess about the forced values, which are set by the current configuration of the giant planets, and thus provide a better window to the history of the solar system than the instantaneous orbital elements. In Figure \ref{fig:proper}, we plot the proper element eccentricities and inclinations of observed KBOs on top of the survival maps of \citet{2005L}, which are formulated in terms of instantaneous eccentricity and inclination. Qualitatively, we see the same features as in Figure \ref{fig:diagnosticall} and Figure \ref{fig:diagnostic}: the cold classicals (red squares) are confined below $e < 0.1$ in the region from 42.5 to 45 AU. Throughout the region, the hot objects (blue triangles) occupy the upper portion of the stability region. 

We repeat the same statistical tests (Table \ref{tab:propert1} and \ref{tab:propert2}) as above for the subset of observed KBOs that have computed proper elements. We use the proper inclinations to classify the objects with $i<2$ as cold. For the second test, for which the sample is probabilistically-selected, we use the inclination distribution from \citet{2011V}, which considers the inclinations with respect to the invariable plane. The results are consistent with the those from the instantaneous orbital elements above. We caution that we are comparing the proper eccentricities of observed objects to a stability map of instantaneous eccentricities. This comparison should be repeated when a stability map formulated in terms of proper elements becomes available.

\begin{table}[hbpt]
\caption{Probability from KS test comparing observed proper eccentricities cold objects with $i<2^\circ$ from Minor Planet Center \citep{2008G,2011V} to ``survival map" distribution.\label{tab:propert1}}
\begin{tabular}{llllll}
				&sample size	&low					&mid					&high\\
$42 < a < 44$ AU	&35			&$10^{-8} $			&$10^{-9}$			&$10^{-9}$\\
$44 < a < 45$ AU	&25 			&$0.002 $				&$0.0006$				&$0.0003$	\\
\end{tabular}

\end{table}

\begin{table}[hbpt]
\caption{Probability from KS test comparing probabilistically-selected observed cold objects (using proper elements) to ``survival map" distribution.\label{tab:propert2}}
\begin{tabular}{llllll}
				&sample size	&low					&mid					&high\\
$42 < a < 44$ AU	\\
				&58			&$10^{-12} $		&$10^{-14}$		&$10^{-14}$\\
				&54			&$10^{-10} $		&$10^{-11}$		&$10^{-12}$\\
				&55			&$10^{-11} $		&$10^{-13}$		&$10^{-13}$\\
				&51			&$10^{-12} $		&$10^{-13}$		&$10^{-14}$\\				
				&49			&$10^{-10} $		&$10^{-11}$		&$10^{-12}$\\												
$44 < a < 45$ AU	
				&40 			&$10^{-5} $		&$10^{-6}$		&$10^{-6}$	\\
				&37 			&$10^{-4} $		&$10^{-4}$		&$10^{-5}$	\\
				&39 			&$10^{-4} $		&$10^{-5}$		&$10^{-5}$	\\
				&37 			&$10^{-4} $		&$10^{-4}$		&$10^{-5}$	\\				
				&36 			&$10^{-4} $		&$10^{-5}$		&$10^{-6}$	\\								
\end{tabular}
\end{table}

\section{Derivation of secular theory}
\label{app:theory}

In Section \ref{app:terms}, we derive additional factors that we include in the secular theory (Section \ref{sec:theory}). In Section \ref{app:planets}, we consider the secular forcing due to other planets besides Neptune and demonstrate that, for KBOs, the secular forcing due to all four planets reduces to the forcing by a precessing Neptune.

\subsection{Derivation of additional terms}
\label{app:terms}

In Section \ref{subsec:refined}, we relegated to the appendix the derivation of several additional terms in the modified secular theory. In Section \ref{subsec:highe}, we derive the factors proportional to $\eN^2$ that appear in the extra factors used in Equation (\ref{eqn:extra}) and (\ref{eqn:extrafactor}) as coefficients to $\eN^2$. In Section \ref{subsec:resterms}, we follow \citet{1989M} to derive resonance correction terms.

\subsubsection{High order eccentricity terms}
\label{subsec:highe}
\begin{figure}[htbp]
\begin{centering}
\includegraphics[width=.5\textwidth]{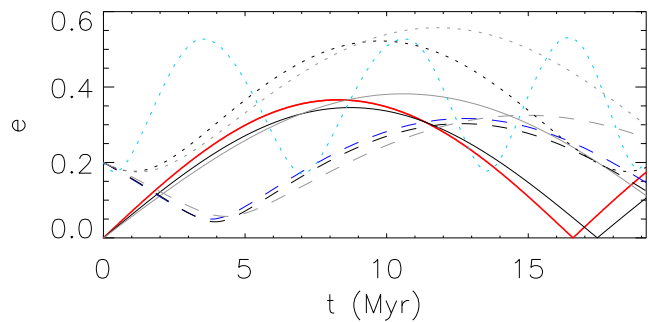}
\caption{Evolution of several example particles illustrating the necessity of the additional $\eN^2$ terms. The colored lines, each representing one of three particles, are the output of a numerical integration in which Neptune has $\eN = 0.3$. Each particle has a different linestyle. The black curve is the analytical model with higher order $\eN$ terms incorporated (Equation \ref{eqn:effects}) while the gray curve is the analytical model that neglects these higher order terms (Equation \ref{eqn:sec}). For each particle, the black curve matches better than the gray curve. When the particle itself has a high eccentricity, depending on its initial phase, the frequency of the analytical model matches well (purple) or is off by a factor of up to two (cyan worst case scenario). The discrepancy between the analytical model and the numerical integration output at some phases for high eccentricity particles is the result of the approximation which is only to lowest order in the particle's eccentricity. Fortunately, the most favorable case for delivering the hot classicals --- an initial periapse that results in the particle's eccentricity sharply decreasing --- is that for which our analytical model performs best. } \label{fig:models}
\end{centering}
\end{figure}

The basic secular theory includes only the lowest order eccentricity terms. However, when Neptune's eccentricity is high, terms containing $\eN^2$ are no longer negligible. Therefore $\gkbo$ and $\bar{e}_{\rm forced}$ must be modified. Here we define the extra terms and factors used in Equation (\ref{eqn:extra}), which come from additional terms in the disturbing function \citep[see Chapter 7 of][for a standard derivation]{2000M}. The disturbing function has the form, up to second order, of:
\begin{eqnarray}
R = n^2 a^2 \frac{\mN}{m_\bigodot} [e^2 (f_2 + f_5 \eN^2 + f_6 e^2) \nonumber \\
+ \cos(\varpi-\varpiN) e \eN (f_{10}  + f_{11} \eN^2 + f{12} e^2) + \cos(2(\varpi-\varpiN)) e^2 \eN^2 f_{17}] \nonumber \\
\end{eqnarray}
Fully incorporating all $e$ and $\eN$ to second order would modify the functional form of the secular theory. However, if we treat $\eN$ as a constant, we can modify the $f_2$ term in the secular forcing frequency $\gkbo $ to $f_2 + f_5 \eN^2$ (Equation \ref{eqn:extra}) and the $f_{10}$ term in the forced eccentricity to $f_{10} + f_{11} \eN^2$ (Equation \ref{eqn:extrafactor}). Because the form of the secular evolution Equation (\ref{eqn:sec}) is derived by differentiating $R$ with respect to the particle's $h$ and $k$, treating $\eN$ as a constant does not modify the form of the secular evolution equations but simply adds extra correction factors. The $f$ factors are defined in Appendix B of \cite{2000M}. The success and necessity of these extra terms is illustrated in Figure \ref{fig:models}.

\subsubsection{Modification near resonances}
\label{subsec:resterms}

Proximity to resonance modifies the secular frequencies. Following \citet{1989M}, we define the resonance-correction terms $C$ and $\epsilon$ in Equations (5) and (6).  \citet{1989M} developed equations for a pair of moons near first order resonance. Instead of a pair of moons, we treat a massless KBO and Neptune, and we include resonances above the first order.

Resonances add additional important terms to the disturbing function, $R$. The factor $C$ in Equation (6) is proportional to the coefficient of the direct part of the disturbing function, with argument $(j+x) \lambda - j \lambda_{\rm N} - x \varpiN$, where $x$ is the order of the resonance. Since the coefficients of this part of the disturbing function $R$ are proportional to $\eN^x$, and since $\dot{h} \propto \frac{\partial R}{\partial h}$ and $\dot{k} \propto \frac{\partial R}{\partial k}$, a factor of $x \eN^{x-1}$ comes in (Equation 6), which was not explicitly included in \citet{1989M} because they treated only the $x=1$ case. The factor $C$ for each order $x$ is tabulated in Table \ref{tab:coeff}. These coefficients were taken from the expansion of the disturbing function in Appendix B of \citet{2000M}.

\begin{table}[hbpt]
\caption{Coefficients for Equation (6). \label{tab:coeff}}
\begin{tabular}{ll}
$x$ & $C$\\
1 & $\frac{1}{2} (-2(j+x) -\alpha \frac{d}{d\alpha})b_{1/2}^{(j+x)}(\alpha)$ \\
2 & $\frac{1}{8} ((-5(j+x)+4(j+x)^2) + (-2+4(j+x))\alpha \frac{d}{d\alpha}$\\
&$+\alpha^2 \frac{d^2}{d\alpha^2}) b_{1/2}^{(j+x)}(\alpha)$ \\
3& $\frac{1}{48} ((-26(j+x)+30(j+x)^2-8(j+x)^3)+ (-9+27(j+x)-12(j+x)^2)\alpha \frac{d}{d\alpha}$\\
&$+(6-6(j+x))\alpha^2 \frac{d^2}{d\alpha^2} - \alpha^3 \frac{d^3}{d\alpha^3} )b_{1/2}^{(j+x)}(\alpha)$ \\
4 & $\frac{1}{384} ((-206(j+x)+283(j+x)^2-120(j+x)^3+16(j+x)^4)+ (-64+236(j+x)$\\
&$-168(j+x)^2+32(j+x)^3)\alpha \frac{d}{d\alpha}$\\
&$+(48-78(j+x)+24(j+x)^2)\alpha^2 \frac{d^2}{d\alpha^2} + (-12+8(j+x)) \alpha^3 \frac{d^3}{d\alpha^3}$\\
&$+ \alpha^4 \frac{d^4}{d\alpha^4} )b_{1/2}^{(j+x)}(\alpha)$ \\
\end{tabular}
\end{table}

The factor $\epsilon$ in Equation (\ref{eqn:extra}) depends on the proximity to resonance. Extending \citet{1989M} to resonances of arbitrary order, we obtain:
\begin{eqnarray}
\omega = j n_N - (j+x) n (1+\frac{\mN}{m_\bigodot}(1+\alpha \frac{d}{d\alpha})b_{1/2}^{(0)} )\nonumber \\
\epsilon = \frac{3}{2} \frac{\mN}{m_\bigodot} (j+x)^2 \frac{1+\frac{\mN}{m_\bigodot} \alpha (1+\frac{7}{3}\alpha \frac{d}{d\alpha}+\frac{2}{3}\alpha^2 \frac{d^2}{d\alpha^2})b_{1/2}^{(0)} }{(\omega/n)^2} \nonumber\\
\end{eqnarray}
where $n_N$ is the mean motion of Neptune and $n$ is the mean motion of the particle.

The $n:1$ resonances have indirect terms not explicitly included in \citet{1989M} (R. Malhotra, private communication). However, the only relevant $n:1$ resonance in the region of the Kuiper Belt we are studying is the $2:1$ resonance and its indirect terms result in expressions that, when incorporated above, are directly proportional to the mass of the KBO and thus assumed to be negligible.

\subsection{Effects of other planets}
\label{app:planets}

In the case of multiple planets, the forced eccentricity of a small body on an external orbit is given by \citep{2000M}:

\begin{eqnarray}
\label{eqn:planets}
h_0= - \sum_{i=1}^N \frac{\nu_i}{A-g_i} \sin (g_i t + \beta_i) \nonumber \\
k_0= - \sum_{i=1}^N \frac{\nu_i}{A-g_i} \cos (g_i t + \beta_i) \nonumber \\
\end{eqnarray}
where
\begin{eqnarray}
\nu_i = \sum_{j=1}^N A_j e_{ji} \nonumber \\
A_j = -n \frac{1}{4} \frac{m_j}{m_\bigodot} \alpha_jb_{3/2}^{(2)}(\alpha_j)\nonumber \\
A = \sum_{j=1}^N n \frac{1}{4}\frac{m_j}{m_c}   \alpha_j b_{3/2}^{(1)}(\alpha_j)\nonumber \\
\end{eqnarray}
where the particle's forced eccentricity $e_{\rm forced} = \sqrt {h_0^2 + k_0^2}$ and $g_i$ and $e_{ji}$ are the eigenfrequencies and eigenvector components of the planetary system. We compare Equation (\ref{eqn:planets}) with Equation (7).  For KBOs in the classical region, Neptune's orbits dominates $A$, the precession rate of the particle's free eccentricity. The precession rate of a particle's free eccentricity due to Neptune alone agrees with the four-planet case to within $30\%$. Thus $ A \approx \gkbo$. Neptune dominates the $\nu_i$ term, and thus the four-planet secular theory reduces to the single-planet secular theory with an extra $g_i = \dotvarpiN$ term for Neptune's precession. Therefore the four planet Equation (\ref{eqn:planets}) reduces to Equation (7). The quantity $\nu_i$ corresponds to $g'_{\rm KBO} \eN (t)$, $g_i$ corresponds to $\dot{\varpi_{\rm N}}$, and $A$ corresponds to $g_{\rm KBO}$. The extra $\sin$ term in Equation (7) is an empirical factor to account for eccentricity damping and is discussed in detail in the main text. This conclusion is consistent with the result of \citet{2008C} that the current forced eccentricities of the KBOs are largely determined by Neptune's orbit.

\end{appendix}
\bibliography{./ms} \bibliographystyle{apj} 

\end{document}